\documentclass[10pt]{iopart}

\usepackage{graphicx}
\usepackage{amsfonts}
\expandafter\let\csname equation*\endcsname\relax
\expandafter\let\csname endequation*\endcsname\relax
\usepackage{amsmath}
\usepackage{enumerate}
\usepackage{comment}
\usepackage{color}
\usepackage{tikz}
\usepackage{citesort}
\usepackage[mathscr]{euscript}

\newcommand{\pprime}{{\prime\prime}}
\newcommand{\one}{{1\!{\rm I}}}

\newcommand{\Z}{\mathbb{Z}}

\newcommand{\bA}{\mathbf{A}}
\newcommand{\bB}{\mathbf{B}}
\newcommand{\berr}{\mathbf{r}}

\newcommand{\bra}{\langle}
\newcommand{\ket}{\rangle}

\newcommand{\bmu}{\boldsymbol{\mu}}

\newcommand{\bphi}{\boldsymbol{\phi}}
\newcommand{\bpsi}{\boldsymbol{\psi}}
\newcommand{\bPhi}{\boldsymbol{\Phi}}
\newcommand{\bPsi}{\boldsymbol{\Psi}}

\newcommand{\bX}{\mathbf{X}}
\newcommand{\bM}{\mathbf{M}}
\newcommand{\bI}{\mathbf{I}}
\newcommand{\bJ}{\mathbf{J}}
\newcommand{\bU}{\mathbf{U}}
\newcommand{\bT}{\mathbf{T}}

\newcommand{\calS}{\mathcal{S}}

\newcommand{\calC}{\mathcal{C}}
\newcommand{\calZ}{\mathcal{Z}}

\newcommand{\bGamma}{\mathbf{\Gamma}}

\newcommand{\calD}{\mathcal{D}}
\renewcommand{\calD}{\mathscr{D}}

\newcommand{\hP}{\hat{P}}
\newcommand{\hW}{\hat{W}}

\newcommand{\bvarphi}{\boldsymbol{\varphi}}
\newcommand{\vsp}{\vspace*{3mm}}

\newcommand{\hrho}{\hat{\varrho}}
\newcommand{\Order}[1]{{\mathcal{O} \left( #1 \right)}}
\newcommand{\s}[1]{\left[ #1 \right]}
\newcommand{\expected}[1]{\left\langle #1 \right\rangle}
\newcommand{\abs}[1]{\left| #1 \right|}
\newcommand{\p}[1]{\left( #1 \right)}

\newcommand{\mue}{{\mu_{\varepsilon}}}

\newcommand{\dmu}{\frac{\rmd}{\rmd\mu}}
\newcommand{\Imag}{\textrm{Im}}
\newcommand{\real}{\textrm{Re}}
\newcommand{\prodmu}{\prod_{\mu}}
\newcommand{\proda}{\prod_{\alpha_\mu =1}^{n_\mu}}
\newcommand{\prodb}{\prod_{\beta_\mu = 1}^{m_\mu}}
\newcommand{\mua}{{\mu , \alpha_\mu}}
\newcommand{\mub}{{\mu , \beta_\mu}}

\newcommand{\vect}[2]{\left( \begin{array}{c}
    #1  \\
    #2 
\end{array}\right)}

\newcommand{\xmu}{x^\star(\mu)}

\begin{document}\vspace*{-30mm}

\title[Imaginary replica analysis of loopy regular random graphs]{Imaginary replica analysis of\\ loopy regular random graphs}

\author{Fabi\'an Aguirre L\'opez$^{1,2}$ and Anthony CC Coolen$^{1,2}$}
\address{$^1$Department of Mathematics, King's College London, The Strand, London WC2R 2LS, United Kingdom\\ 
$^2$Institute for Mathematical and Molecular Biomedicine, King's College London, Hodgkin Building, London SE1 1UL, United Kingdom
}

\ead{fabian.aguirre\_lopez@kcl.ac.uk, ton.coolen@kcl.ac.uk}

\begin{abstract}
We present an analytical approach for describing spectrally constrained maximum entropy ensembles of  finitely connected regular loopy graphs, valid in the regime of weak loop-loop interactions. We derive an expression for the leading two orders of the expected eigenvalue spectrum, through the use of infinitely many replica indices taking imaginary values. We apply the method to models in which the spectral constraint reduces to a soft constraint on the number of triangles, which exhibit `shattering' transitions to phases with extensively many disconnected cliques, to models with controlled numbers of triangles and squares, and to models where the spectral constraint reduces to a count of the number of adjacency matrix eigenvalues in a given interval.  Our predictions are supported by MCMC  simulations  based on edge swaps with nontrivial acceptance probabilities.
\end{abstract}


\section{Introduction}

A huge amount of scientific work has been devoted in recent decades to the  study of random graphs, motivated partly by their interesting mathematical properties  \cite{bollobas2001random} and  partly because of their frequent and fruitful use in the modelling of complex systems \cite{newman2010networks}. The most famous random graph ensembles are probably the Erd\"os-Renyi model (ER) \cite{solomonoff1951connectivity,erdos1959random,erdos1960evolution} and the configuration model (CM) \cite{bender1978asymptotic}. These are popular because models of interacting systems defined on their typical graph instances  can often be solved analytically, and  because sampling graphs from these ensembles is easy. Both  properties derive mainly from the fact that ER and CM graphs are typically locally tree like, which enables the application of many relatively simple mathematical and numerical approaches. In fact, nearly all mathematical and computational techniques currently available for analysing processes on large graphs (cavity methods, belief propagation and other message passing algorithms, generating functional analysis, conventional replica methods, etc)  rely explicitly or implicitly on being able to neglect the presence of short loops, or on being able to treat such loops as perturbations of a fundamentally tree-like architecture.

 Ironically, their built-in locally tree like topology makes both ER and CM ensembles, and  their numerous tree-like variations such as  scale-free network models,  also rather unsuitable as  models for real networks.  Simplicity comes at a nontrivial cost. Real networks are usually not tree like, but have many short loops, i.e. closed non backtracking paths \cite{newman2002random}. Moreover, their significant `loopiness' cannot be ignored, since it has profound functional implications\footnote{For instance, extensive loopiness is the main topological difference between the trivially solved ferromagnets on Bethe lattices or on random regular graphs with degree six, and the as yet unsolved three-dimensional Ising model.} .  Hence there is a clear need for simple loopy graph ensembles, and for new mathematical tools with which to analyse them.  

Studying random graphs is also important in the area of  null models for real observed networks. Null models of networks are unbiased random graph ensembles describing graphs that share with the networks the values of a given set of observables, i.e. maximum entropy ensembles \cite{park2004statistical,annibale2017generating} subject to specific topological constraints. The constraints can be imposed as `soft' conditions, where observables are matched on average, or as `hard' conditions, where observables are matched by every graph in the ensemble. Realistic null models for graphs must be sparse and loopy. The  simplest approach would be to control the number of edges and the number of triangles via soft constraints, 
leading to e.g.  exponential ensemble of \cite{strauss1986general}. Unfortunately, upon varying the control parameters this ensemble was found to switch between very weak clustering and dominance by dense graphs \cite{jonasson1999random,chatterjee2013estimating,yin2016detailed}. See also \cite{burda2004network}. The alternative is to impose both the number of single edges and of edges in a triangle around a node via hard constraints  \cite{newman2009random}, but these graphs do not lend themselves as easily to tractable solutions, see e.g. \cite{peron2018spectra,herrero2015ising,herrero2019self}.

In the present paper we focus only maximum entropy ensembles of  $q$-regular graphs, with soft-constrained adjacency matrix eigenvalue spectra to introduce `loopiness'. We use the imaginary replica approach to obtain analytical results on the dependence of the spectrum on the functional Lagrange parameter,  following  \cite{CoolenLoopy}, and we build on recent studies such as 
\cite{cavagna2000index,metz2015index,metz2016large,metz2017replica,castillo2018large,castillo2018theory}. The functional nature of the constraint in our ensemble allows us to bias the number of eigenvalues in \emph{all} infinitesimal intervals simultaneously, while recovering the average spectral density through a simple functional derivative. We compute  analytically  in leading two orders in the system size  the expected spectrum, a calculation found to  take its most natural form using Chebyshev polynomials, and is valid in the regime where the loops are still sufficiently rare to prevent loop-loop interactions from becoming relevant. 
The simplest nontrivial members of our family of spectrally constrained maximum entropy ensembles are those where the spectral constraint reduces to  a 
constraint on the expected number of triangles (equivalently, on the average clustering coefficient), with a single Lagrange parameter $\alpha$. We show that these models always exhibit transitions into a phase where the graphs shatter into an extensive number of disconnected cliques of $q+1$ nodes,  analogous to the condensation transition in the model in \cite{strauss1986general}. We are able to present a description of the ensemble for all values of $\alpha$ for large enough $N$. We also show how our general theory can be applied to compute spectra for other ensembles with more complicated Lagrange parameters. 

Our paper is organised as follows. In section 2 we give the relevant definitions of our spectrally constrained ensembles. Section 3 is devoted to the derivation of the generating function for the adjacency matrix spectrum. The theory is then applied in section 4, first to existing models (recovering known results as a test), followed by applications to other loopy graph ensembles. Technical details are often delegated to Appendices, to improve the flow of the paper. We end with a discussion of the results obtained, and an outlook on future work.

\section{Definitions}

We study ensembles of simple nondirected $N$-node regular graphs with degree $q$. Each graph is defined by its symmetric $N\times N$ adjacency matrix  $\bA$, with entries $A_{ij}=1$ if nodes $i$ and $j$ are connected,  and $A_{ij}=0$ otherwise ($A_{ii}=0$ for all $i$). 
A path of length $\ell$ on a graph is a sequence of $\ell+1$ pairwise connected nodes. A closed path  starts and ends in the same node. Loops are closed paths without repetition of nodes, except for the first and the last node. The formula for the total number of closed paths of a given length in a graph $\bA$ is, apart from a simple overcounting factor, 
\begin{eqnarray}
    \Tr(\bA^\ell) &=& \textrm{\# of closed paths of length }\ell.
\end{eqnarray}
It  follows from  the relation  $\Tr(\bA^\ell)  = N\int\! \rmd\mu\,\varrho(\mu|\bA) \mu^\ell$. 
that controlling the numbers of closed paths of all lengths $\ell$ in random graphs is equivalent to controlling the moments of the spectral density $ \varrho(\mu|\bA) $ of $\bA$,
\begin{eqnarray}
    \varrho(\mu|\bA) = \frac{1}{N}\sum_{i=1}^N\delta(\mu-\mu_i(\bA)),
\end{eqnarray}
where $\mu_i(\bA)$  is the $i$-th eigenvalue of  $\bA$. 

In exponential spectrally constrained ensembles \cite{CoolenLoopy}, the graph probabilities $p(\bA)$ on the set $G$ of simple nondirected $N$-node graphs are determined by maximising the Shannon entropy $S[p]=\sum_{\bA\in G}p(\bA)\log p(\bA)$, subject to prescribed values of all degrees and a prescribed expectation value of the spectral density. For $q$-regular random graphs this gives
\begin{eqnarray}
    p(\bA) &=& \frac{\rme^{N\int\!\rmd\mu~\hrho(\mu)\varrho(\mu|\bA)}}{Z[\hrho]}\prod_{i=1}^N \delta_{q,\sum_{j}A_{ij}},
     \label{eq:spectral_ensemble}
    \\
    Z[\hat{\varrho}]&=&\sum_{\bA\in G}\rme^{N\int\!\rmd\mu~\hrho(\mu)\varrho(\mu|\bA)}\prod_{i=1}^N \delta_{q,\sum_{j}A_{ij}}.
       \end{eqnarray}
    Here $\hrho(\mu)$ is a functional Lagrange multiplier. 
   By construction,  (\ref{eq:spectral_ensemble}) defines the most unbiased ensemble of $q$-regular nondirected graphs with a prescribed adjacency matrix spectrum.   We write averages over  (\ref{eq:spectral_ensemble}) as  $\bra f(\bA)\ket=\sum_{\bA\in G}p(\bA)f(\bA)$.
    The expected eigenvalue density $\varrho(\mu)=\bra \varrho(\mu|\bA)\ket$ can be obtained from a generating function $\phi[\hrho]$:
    \begin{eqnarray}
 \varrho(\mu)&=&
 \frac{\delta \phi[\hrho]}{\delta \hrho(\mu)},~~~~~~\phi[\hat{\varrho}]=N^{-1}\log Z[\hat{\varrho}].
 \label{eq:spectralEnsemble}
\end{eqnarray}
Our main interest is in finding an analytical expression for the expected density $\varrho(\mu)$  in terms of the functional Lagrange parameter $\hat{\varrho}(\mu)$.

For the simple choice $\hrho(\mu) = \alpha\mu^3$ we recover from  (\ref{eq:spectral_ensemble}) the model of \cite{lopez2018exactly}, in which the number of loops of length three (i.e. of triangles) is constrained:
\begin{eqnarray}
\hspace*{-10mm}
    p(\bA) &=& \frac{\rme^{\alpha \Tr (\bA^3) }}{Z(\alpha)}\prod_{i=1}^N\delta_{q,\sum_j A_{ij}}, ~~~~~~Z(\alpha)=\sum_{\bA\in G}\rme^{\alpha \Tr (\bA^3) }\prod_{i=1}^N\delta_{q,\sum_j A_{ij}}.
    \label{eq:ensemble}
    \end{eqnarray}
 The number of triangles $n_{\triangle}(\bA)$ differs from  the trace  only by an overcounting factor, viz. $n_{\triangle}(\bA)=\frac{1}{6}\Tr(\bA^3) $.
 The generating function associated with (\ref{eq:ensemble}) is $\phi(\alpha) = N^{-1}\log Z(\alpha) $.   From $m(\alpha)=\partial \phi(\alpha)/\partial\alpha =N^{-1}\bra {\rm Tr}(\bA^3)\ket$ follows the average clustering coefficient  
   $\bra C(\bA)\ket=m(\alpha)/q(q-1)$. For $q=2$ one can compute $\phi(\alpha)$ using standard combinatorics \cite{lopez2018exactly}, for both $\alpha=\Order{N^0}$ and  $\alpha=\Order{\log N}$. 
For arbitrary $q$, the expected loop density $m(\alpha)$ always vanishes for $N\to\infty$ if $\alpha=\Order{1}$. As in \cite{lopez2018exactly}, one could rescale $\alpha$ with a factor $\log N$. While this could give an asymptotic theory with finite loop densities, we will find that it would not be a useful network model for applications.

\section{Evaluation of the generating function}

\subsection{Imaginary replica approach}

We note that all $q$-regular graphs of size $N$ have identical probabilities in an Erd\"os-Renyi (ER) ensemble with  average degree $q$, 
\begin{eqnarray}
\hspace*{-20mm}
    p_{\rm ER}(\bA) = \Big(\frac{q}{N}\Big)^{\!\frac{Nq}{2}} \Big(\frac{N-q}{N}\Big)^{\!\frac{N}{2}(N-1-q)}~~~{\rm for~all~}\bA\!\in\! G~~{\rm with}~~\prod_{i=1}^N\delta_{q,\sum_j A_{ij}}=1.
\end{eqnarray}
Hence we can rewrite  (\ref{eq:spectralEnsemble}) as an average over this ER ensemble:
\begin{eqnarray}
   \phi[\hrho] &=& \frac{1}{N}\log \Big\bra \rme^{N\int\!\rmd\mu~\hrho(\mu)\varrho(\mu|\bA)}\prod_{i=1}^N \delta_{q,\sum_j A_{ij}}\Big\ket_{\rm ER}
 +{\rm constant}.
 \label{eq:generator_in_ER}
\end{eqnarray}
Upon using the  Edwards-Jones formula \cite{edwards1976eigenvalue} for $\varrho(\mu|\bA)$, writing the integral over eigenvalues as $\int\!\rmd\mu\ldots =\lim_{\Delta\to 0}\Delta\sum_\mu\ldots$, and after some modest manipulations, 
the key quantity in this expression can be written as follows, with infinitely many imaginary replicas (two   for each eigenvalue$\mu$, $n_\mu$ and $m_\mu$):
\begin{eqnarray}
\hspace*{-10mm}
    \rme^{N\int\!\rmd\mu~\hrho(\mu) \varrho(\mu|\bA)} &=& \lim_{\Delta\to 0}\lim_{n_\mu \to \rmi\frac{\Delta}{\pi}\hrho^\prime(\mu)}\lim_{m_\mu\to -n_\mu}\prod_{\mu}Z(\mue|\bA)^{n_{\mu}}\overline{Z(\mue|\bA)}^{m_\mu}.
    \label{eq:spectrum_formula}
\end{eqnarray}
Here $\mue = \mu + \rmi\varepsilon$ and 
\begin{eqnarray}
   Z(\mue|\bA) &=& \int\prod_{i=1}^N \rmd\phi^i  \exp\Big[\!-\!\frac{\rmi}{2}\sum_{i j }\phi^i(A_{ij}\!-\!\mue \delta_{ij})\phi^j\Big].
   \label{eq:originalZ}
\end{eqnarray}
One initially takes $n_\mu , m_\mu \in \mathbb{N}$, in order to perform the calculation, followed by analytic continuation to the relevant imaginary values. In its above form, (\ref{eq:spectrum_formula}) appeared first in \cite{roberts,CoolenLoopy}, but similar formulae involving limits of replica dimensions to non-zero values have been introduced  in different  contexts, in particular when counting the number of eigenvalues in certain intervals for random matrix ensembles, see e.g. \cite{cavagna2000index,metz2015index,metz2016large,metz2017level,metz2017replica,castillo2018large,castillo2018theory}.
We can combine the integrals in (\ref{eq:spectrum_formula}) as follows:
\begin{eqnarray}
&&\hspace*{-22mm}
    \prodmu Z(\mue|\bA)^{n_{\mu}}\overline{Z(\mue|\bA)}^{m_\mu}  =  \int\!\rmd[\bPhi,\bPsi] \exp\Big(\!-\frac{\rmi}{2}\sum_{\mua}\sum_{ij}\phi^i_\mua \phi^j_\mua (A_{ij} \!-\!\mue \delta_{ij})\Big)
    \nonumber\\
    && \hspace*{25mm} \times \exp\Big(\frac{\rmi}{2}\sum_\mub \sum_{ij} \psi^i_\mub \psi^j_\mub (A_{ij} \!- \overline\mue \delta_{ij})\Big),
\end{eqnarray}
where $d[\bPhi,\bPsi] \!=\! \prod_{i=1}^N \prodmu \big[\big(\proda\! \rmd\bphi^i_\mua\big)\big( \prodb \!\rmd \psi^i_\mub\big)\big]$.  To simplify our notation we introduce the vector  $\bvarphi\in\mathbb{R}^{d}$, where $d = \sum_{\mu}n_\mu \!+\!\sum_{\mu}m_\mu$, with entries  
\begin{eqnarray}
    \bvarphi = \vect{ \{\phi_\mua\}}{\{\psi_\mub\}},
\end{eqnarray}
 the dot product 
\begin{eqnarray}
\label{eq:dotproduct}
    \bvarphi\cdot\bvarphi^\prime = \sum_\mua \phi_\mua \phi^\prime_\mua -\sum_\mub \psi_\mub \psi^\prime_\mub,
\end{eqnarray}
and  the following $d\!\times\! d$ diagonal matrix $\bmu$, in which $\mu\in\{\mu_1,\ldots,\mu_M\}$ and $\bI_n$ denotes the $n$ dimensional identity matrix:
\begin{eqnarray}
    \bmu= \left(\begin{array}{cccccc}
    {\mu_1}_\varepsilon \bI_{n_{\mu_1}}     &\mathbf{0}&&\dots&&\mathbf{0}\\
    \mathbf{0}    & \ddots &\mathbf{0}&&&\\
    & \mathbf{0} &{\mu_M}_\varepsilon \bI_{n_{\mu_M}} &\mathbf{0}&&\vdots\\
    \vdots&&\mathbf{0}&\overline{{\mu_1}_\varepsilon}\bI_{m_{\mu_1}}&\mathbf{0}&\\
    &&&\mathbf{0}&\ddots&\mathbf{0}\\
    \mathbf{0}&&\dots&&\mathbf{0}&\overline{{\mu_M}_\varepsilon}\bI_{m_{\mu_M}}
    \end{array}\right).~
\end{eqnarray}
Here $M$ denotes the number of $\mu$-values in the discretized eigenvalue integral, so $M\to\infty$ when we take the limit $\Delta\to 0$.
 We finally introduce the two shorthands
\begin{eqnarray}
    \nu(\bvarphi) = \rme^{\frac{1}{2}\rmi\bvarphi\cdot\bmu\bvarphi},~~~~~~
    \lim =\lim_{\Delta \to 0} \lim_{n_\mu \to \rmi\frac{\Delta}{\pi}\hrho'(\mu)} \lim_{m_\mu \to -n_\mu}.
    \label{eq:imaginary_limits}
\end{eqnarray}
The above definitions, together with the integral form 
 of the Kronecker delta, $\delta_{nm} =(2\pi)^{-1}\! \int_{-\pi}^\pi\!\rmd\omega~\exp[\rmi \omega (n-m)]$, enable us to compute the generating function  (\ref{eq:generator_in_ER}) following \cite{CoolenLoopy}, which in turn is reminiscent of previous spectral calculations for sparse random graphs \cite{kuhn2008spectra,metz2014finite}. Upon dropping the irrelevant constant in (\ref{eq:generator_in_ER}) we get
\begin{eqnarray}
\hspace*{-20mm}
    \phi[\hrho] &=& \lim \frac{1}{N} \log \int\prod_{i=1}^N\Big[ \rmd\bvarphi^i~ \nu(\bvarphi^i) \frac{\rmd\omega_i}{2\pi} \, \rme^{\rmi \omega_i q} \Big]\expected{\rme^{-\rmi \sum_{i<j} A_{ij} [\bvarphi^i\cdot\bvarphi^j  + \omega_i + \omega _j]}}_{\rm ER}
    \nonumber\\
    \hspace*{-20mm}
  &=&\lim\frac{1}{N}\log \int \prod_{i=1}^N\Big[ \rmd\bvarphi^i~ \nu(\bvarphi^i) \frac{\rmd\omega_i}{2\pi} \, \rme^{\rmi \omega_i q}\Big]\rme^{\sum_{i < j}\log\big[ 1 + \frac{q}{N}\big(\rme^{-\rmi [\bvarphi^i \cdot \bvarphi^j + \omega_i +\omega_j]} - 1\big)\big]}.
\end{eqnarray}
Since we intend to compute finite size spectrum fluctuations, in expanding the logarithm for large $N$ 
we keep both the $\Order{N}$ and $\Order{1}$ terms in the exponent:
\begin{eqnarray}
\hspace*{-21mm}
    \phi[\hrho] &=&
     -\frac{q}{2}(1\!+\! \frac{q\!-\!2}{2N})
   + \lim\frac{1}{N}\log \!\int \!\prod_{i=1}^N\Big[ \rmd\bvarphi^i \nu(\bvarphi^i) \frac{\rmd\omega_i}{2\pi} \, \rme^{\rmi \omega_i q
     - \frac{q}{2N}\rme^{-\rmi [\bvarphi^i \cdot \bvarphi^i+ 2\omega_i ]} }
     \Big]
      \\
     \hspace*{-25mm}
&&
\hspace*{-3mm}
\times \exp\Big[
 \frac{q}{2N}(1\!+\!\frac{q}{N})\sum_{i j}\rme^{-\rmi [\bvarphi^i \cdot \bvarphi^j + \omega_i +\omega_j]} 
 \!
-\! \frac{q^2}{4N^2}\sum_{i j}\rme^{-2\rmi [\bvarphi^i \cdot \bvarphi^j + \omega_i +\omega_j]}  
\Big]+{\cal O}(\frac{1}{N^{2}}).
   \nonumber
\end{eqnarray}
We now introduce the order parameter
\begin{eqnarray}
    P(\bvarphi, \omega) = \frac{1}{N}\sum_{i=1}^N \delta(\bvarphi - \bvarphi^i) \delta(\omega- \omega_i)
\end{eqnarray}
We enforce it by inserting the following functional integral, obtained by writing delta functions for each $(\bvarphi,\omega)$ in integral representation, and with the usual path integral measure $\calD P = \prod_{\bphi}\prod_{\omega}[ \rmd P(\bphi,\omega) \sqrt{N\Delta_{\varphi}\Delta_\omega /2\pi)}]$ (where $\Delta_{\varphi},\Delta_\omega\to 0$):
\begin{eqnarray}
\hspace*{-10mm}
    1 = \int \calD P \calD \hP  ~\rme^{N\rmi \int\!\rmd\bvarphi\rmd\omega P(\bvarphi,\omega)\hP(\bvarphi,\omega)- \rmi \int\! \rmd \bvarphi \rmd \omega \hP(\bvarphi,\omega)\sum_i \delta(\bvarphi -  \bvarphi^i) \delta(\omega-\omega_i)}.
\end{eqnarray}
The result  is:
\begin{eqnarray}
\hspace*{-12mm}
    \phi[\hrho] &= & \lim \frac{1}{N}\log \int\! \calD P \calD \hP~\rme^{N\calS [P,\hP]} +{\cal O}(\frac{1}{N^{2}}),
    \label{eq:generator_in_nasty_integral}
    \\ \hspace*{-12mm}
    S[P,\hP] &= & 
      -\frac{q}{2}(1\!+\! \frac{q\!-\!2}{2N})
      +\rmi \!\int\!\rmd\bvarphi\rmd\omega ~P(\bvarphi,\omega)\hP(\bvarphi,\omega)
            \nonumber
            \\
      \hspace*{-12mm}
      &&
      + \frac{q}{2}(1\!+\!\frac{q}{N})
\! \int\!\rmd\bvarphi\rmd\bvarphi\rmd\omega\rmd\omega^\prime~P(\bvarphi,\omega)P(\bvarphi^\prime,\omega^\prime)
 \rme^{-\rmi \bvarphi \cdot \bvarphi^\prime - \rmi\omega -\rmi\omega^\prime} 
 \nonumber
 \\
 \hspace*{-12mm} &&
-\frac{q^2}{4N}\! \int\!\rmd\bvarphi\rmd\bvarphi\rmd\omega\rmd\omega^\prime~
P(\bvarphi,\omega)P(\bvarphi^\prime,\omega^\prime)
\rme^{-2\rmi \bvarphi \cdot \bvarphi^\prime - 2\rmi\omega -2\rmi\omega^\prime}  
\label{eq:nasty_functional_integral}
      \\
    \hspace*{-12mm}  &&
     - \frac{q}{2N}\!
     \int\!\rmd\bvarphi\rmd\omega~P(\bvarphi,\omega)
     \rme^{-\rmi \bvarphi \cdot \bvarphi- 2\rmi\omega} 
   + \log \!\int \! \frac{\rmd\omega \rmd\bvarphi}{2\pi} \nu(\bvarphi) 
   \rme^{\rmi \omega q- \rmi  \hP(\bvarphi,\omega)}.
 \nonumber
    \end{eqnarray}
It was shown in \cite{metz2014finite} how this type of integral can be reduced to an integral over a single functional variable. In  \ref{appendix:changeVariable} we work out the details, leading to 
\begin{eqnarray}
\hspace*{-18mm}
    \phi[\hrho] &=& \lim\Big\{ \log \int\!\rmd\bvarphi~\nu(\bvarphi)\Big[\int\!\rmd\bvarphi^\prime~U_1(\bvarphi,\bvarphi^\prime)W_0(\bvarphi^\prime)\Big]^q\!+\frac{1}{2N}\sum_{\ell=3}^\infty \frac{{\rm Tr}(\bT^\ell)}{\ell}\Big\},
    \label{eq:result_before_RS}
    \end{eqnarray}
    in which $U_1(\bvarphi,\bvarphi^\prime)=\rme^{-\rmi\bvarphi\cdot\bvarphi^\prime}$, and the function $W_0(\bvarphi)$ is to be solved from 
    \begin{eqnarray}
    W_0(\bvarphi)&=& \frac{\nu(\bvarphi)}{Z_q}\Big[\int\!\rmd\bvarphi^\prime~U_1(\bvarphi,\bvarphi^\prime)W_0(\bvarphi^\prime)\Big]^{q-1},
    \label{eq:spe_RSB}
    \\
    Z_q&=& \int\!\rmd\bvarphi ~\nu(\bvarphi)\Big[\int\!\rmd\bvarphi^\prime~U_1(\bvarphi,\bvarphi^\prime)W_0(\bvarphi^\prime)\Big]^{q},
    \end{eqnarray}
    and
    \begin{eqnarray}
    \hspace*{-15mm}
    T(\bvarphi,\bvarphi^\prime)&=& (q\!-\!1)r[W_0(\bvarphi)]U_1(\bvarphi,\bvarphi^\prime)-q W_0(\bvarphi)\int\!\rmd\bpsi~U_1(\bvarphi^\prime,\bpsi)W_0(\bpsi),
    \label{eq:Tkernel}
    \\
     \hspace*{-15mm}
     r[W_0(\bvarphi)]&=& \frac{\nu(\bvarphi)}{Z_q}
  \Big[\int\!\rmd\bvarphi^\prime ~ U_1(\bvarphi,\bvarphi^\prime) W_0(\bvarphi^\prime)\Big]^{q-2}.
    \end{eqnarray}
   Finally, following  \cite{lucibello2014finite} we may use the following identity\footnote{The proof of this interesting identify follows directly from  the operator properties $\bM\bB=\bB\bM=\bB^2=\one$, where $M(\bvarphi,\bvarphi^\prime)=r[W_0(\bvarphi)]U_1(\bvarphi,\bvarphi^\prime)$ and $B(\bvarphi,\bvarphi^\prime)=W_0(\bvarphi)\int\!\rmd\bpsi~U_1(\bvarphi^\prime,\bpsi)W_0(\bpsi)$.}:
   \begin{eqnarray}
   {\rm Tr}(\bT^\ell) = (q\!-\!1)^{\ell}[{\rm Tr}(\bM^\ell) \!-\! 1] + (-1)^\ell 
   \end{eqnarray}
  with $M(\bvarphi,\bvarphi^\prime)=r[W_0(\bvarphi)]U_1(\bvarphi,\bvarphi^\prime)$, to simplify 
    (\ref{eq:result_before_RS}) modulo an additive constant to
    \begin{eqnarray}
    \phi[\hrho] &=& \lim\Big\{ \log Z_q+\sum_{\ell=3}^\infty \frac{(q\!-\!1)^{\ell}}{2N\ell}
    {\rm Tr}(\bM^\ell) \Big\}.
    \label{eq:result_before_RS2}
    \end{eqnarray}
Expression  (\ref{eq:result_before_RS2}) was originally presented in \cite{metz2014finite}, and we have indeed chosen our notation at the start deliberately to emphasize and exploit the similarity. However, although identical in structure, the present formula (\ref{eq:result_before_RS2})  differs from the one in \cite{metz2014finite}  in the underlying definitions of the fields $\bvarphi$, the dot product $\bvarphi\cdot\bvarphi^\prime$ and the function $\nu(\bvarphi)$, which here all  involve  the full eigenvalue spectrum and describe our present controlled non-uniform measures over the space of graphs.

\subsection{Replica symmetric solution}

In order to continue, we assume that the order parameter $W_0(\bvarphi)$ is replica symmetric (RS), i.e. invariant under all permutations of all replicas (noting that in the present problem we have a separate replica index for each eigenvalue $\mu$, and that $W_0(\bvarphi)$ is not a normalized distribution). We write $W_0(\bvarphi)$ as a superposition of zero mean complex Gaussian distributions:
\begin{eqnarray}
    \label{eq:repsym}
    W_0(\bvarphi) &= & \calC \int\! \rmd\bX~ W(\bX) \frac{1}{Z(\bX)}\rme^{-\frac{1}{2}\rmi\bvarphi\cdot \bX\bvarphi},\\
    Z(\bX) &= & \int\!\rmd\bvarphi~\rme^{-\frac{1}{2}\rmi\bvarphi\cdot\bX\bvarphi} =
    \prod_{\mu}\Big(\frac{2\pi}{\rmi x(\mu)}\Big)^{\!\!\frac{n_\mu}{2} } \Big(\overline{\frac{2\pi}{\rmi x(\mu)}}\Big)^{\frac{\!\!m_{\mu}}{2}},
    \label{eq:RS_order_parameter}
\end{eqnarray} 
where $\bX\in\mathbb{R}^{d \times d}$ is a diagonal matrix with the following structure:
\begin{eqnarray}
\hspace*{-20mm}
    \bX = \left(\begin{array}{cccccc}
    x\p{\mu_1} \bI_{n_{\mu_1}}     &\mathbf{0}&&\dots&&\mathbf{0}\\
    \mathbf{0}    & \ddots &\mathbf{0}&&&\\
    & \mathbf{0} &x\p{\mu_M} \bI_{n_{\mu_M}} &\mathbf{0}&&\vdots\\
    \vdots&&\mathbf{0}&\overline{x\p{\mu_1}}\bI_{m_{\mu_1}}&\mathbf{0}&\\
    &&&\mathbf{0}&\ddots&\mathbf{0}\\
    \mathbf{0}&&\dots&&\mathbf{0}&\overline{x\p{\mu_M}}\bI_{m_{\mu_M}}
    \end{array}\right).
\end{eqnarray}
Expression (\ref{eq:RS_order_parameter}) 
 is indeed invariant under all permutations of replica indices with any fixed value of $\mu$, that is $\{\phi_{\mu,1},\ldots,\phi_{\mu,n_\mu}\}$ and $\{\psi_{\mu,1},\ldots,\psi_{\mu,m_\mu}\}$. The new RS order parameter is the distribution $W(\bX)$, where each $\bX$ is specified by $M$ complex numbers $x(\mu)$. Hence the integration in  (\ref{eq:dotproduct})   is over the real and imaginary part of each $x(\mu)$, so $\rmd\bX = \prodmu \rmd\,\real[x(\mu)] \rmd\,\Imag[x(\mu)]$. For (\ref{eq:repsym}) to be well defined, we must restrict all $x(\mu)$ to have $\Imag\; x(\mu)<0$. We may assume that $\int\!\rmd\bX~ W(\bX) =1$, since possible non-normalization of $W_0$ is reflected in the inclusion in (\ref{eq:RS_order_parameter})  of a constant $\calC$.

We note that for the present definition  (\ref{eq:dotproduct})  of the dot product, the following identity is still valid.
\begin{eqnarray}
    \int\!\rmd\bvarphi^\prime~\rme^{-\rmi\bvarphi\cdot\bvarphi^\prime-\frac{1}{2}\rmi\bvarphi^\prime\cdot \bX \bvarphi^\prime} = Z(\bX)\rme^{\frac{1}{2}\rmi\bvarphi\cdot\bX^{-1} \bvarphi}.
\end{eqnarray}
Insertion of our RS ansatz   (\ref{eq:RS_order_parameter}) into the full order parameter equation (\ref{eq:spe_RSB}) shows, using the above identity,  that the RS ansatz indeed gives a solution of (\ref{eq:spe_RSB}), provided the RS order parameter satisfies
\begin{eqnarray}
    W(\bX) &=& \frac{Z(\bX) }{\calZ_{q-1}}\int\Big(\prod_{k=1}^{q-1}\rmd\bX_k W(\bX_k)\Big) ~ \delta\Big(\bX + \bmu + \sum_{k=1}^{q-1}\bX_k^{-1}\Big),
    \label{eq:RS_W}
    \\
    \calZ_q &= & \int\!\rmd\bX \Big(\prod_{k=1}^q \rmd\bX_k W(\bX_k) \Big)~ Z(\bX)~\delta\Big(\bX + \bmu + \sum_{k=1}^q \bX_k^{-1}\Big),
    \label{eq:RS_Z}
    \\
    \calC^2 &= & \calZ_{q-1}/{\calZ_q}.
    \label{eq:bigC}
\end{eqnarray}
We similarly derive $Z_q=C^q\calZ_q$, giving in combination with (\ref{eq:bigC}):
\begin{eqnarray}
\log Z_q&=&\frac{1}{2}q\log  \calZ_{q-1}-\frac{1}{2}(q\!-\!2)\log \calZ_q.
\label{eq:logZ}
\end{eqnarray}

The above RS order parameter equations (\ref{eq:RS_W},\ref{eq:RS_Z})  have one specific  simple solution, namely the delta distribution $W(\bX) = \delta(\bX - \bX^\star)$, in which the entries of $\bX^\star$ satisfy
\begin{eqnarray}
    \xmu = -\mue - \frac{q-1}{\xmu}.
    \label{eq:xstar}
\end{eqnarray}
Of the two possible solutions of this equation we must choose the one with $\Imag\;x(\mu)<0$:
\begin{eqnarray}
\label{solutionXstar}
    \xmu &= -\frac{1}{2}\mue-\frac{1}{2}\rmi\sqrt{4(q\!-\!1)-\mue^2}.
    \label{eq:xstarmu}
\end{eqnarray}
For this special RS solution we have 
\begin{eqnarray}
 W_0(\bvarphi)&=& \frac{\calC}{Z(\bX^\star)}\rme^{-\frac{1}{2}\rmi\bvarphi\cdot \bX^\star\bvarphi},
\\
  \calZ_{q-1}&=& Z(\bX^\star),
  \label{eq:calZ_q-1}
  \\ 
   \calZ_q &= & \int\!\rmd\bX ~ Z(\bX)~\delta\Big(\bX \!+ \!\bmu \!+ \!q (\bX^\star)^{-1}\Big),\\
   \nonumber &=& Z(\bX^\star\!-(\bX^\star)^{-1}),
     \label{eq:calZ_q}
\end{eqnarray}
and hence also $\calC^2=Z(\bX^\star)/Z(\bX^\star\!-(\bX^\star)^{-1})$. 
The kernel $\bM$ which appears in the generating function  (\ref{eq:result_before_RS2}) will now have the following entries:
\begin{eqnarray}
M(\bvarphi,\bvarphi^\prime)&=& \calZ^{-1}_{q-1}~\rme^{\frac{1}{2}\rmi\bvarphi\cdot[\bmu+(q-2) (\bX^\star)^{-1}]\bvarphi-\rmi\bvarphi\cdot\bvarphi^\prime}.
\label{eq:RS_M}
\end{eqnarray}
In \ref{app:traces} we work out the traces ${\rm Tr}(\bM^\ell)$ for the RS solution, and find
\begin{eqnarray}
{\rm Tr}(\bM^\ell)
     &=&  \calZ^{-\ell}_{q-1} \prodmu \Big[Z(\mue|\bA^\star_{\ell,\mu})^{n_\mu}\overline{Z(\mue|\bA^\star_{\ell,\mu})}^{m_\mu}\Big].
\end{eqnarray}
Where $Z(\mue|\bA^\star_{\ell,\mu})$ denotes the original complex Gaussian integral defined in (\ref{eq:originalZ}), and  $\bA^\star_{\ell,\mu}$ is now the $\ell\times\ell$ adjacency matrix of a loop of length $\ell$ in the presence of a complex field acting on  the diagonal, of value $(2\!-\!q)/\xmu$:
\begin{eqnarray}
    \p{\bA^\star_{\ell,\mu}}_{kk^\prime} &=& \delta_{k,k^\prime\!+1} +\delta_{k,k^\prime\!-1} +\frac{2\!-\!q}{\xmu}\delta_{kk^\prime}~~~~~({\rm with}~k~{\rm mod}~\ell).
    \label{eq:TraceDone}
\end{eqnarray}
Substituting (\ref{eq:logZ}) and (\ref{eq:TraceDone})  into expression (\ref{eq:result_before_RS2}) for the generating function, folowed by using formulae  (\ref{eq:calZ_q-1},\ref{eq:calZ_q}) for the constants $\calZ_{q-1}$ and $\calZ_q$, then gives 
   \begin{eqnarray}
    \phi[\hrho] &=& \lim\Big\{ \frac{1}{2}q\log Z(\bX^\star)-\frac{1}{2}(q\!-\!2)\log Z(\bX^\star\!-(\bX^\star)^{-1})
    \nonumber
    \\
    &&+\sum_{\ell=3}^\infty \frac{(q\!-\!1)^{\ell}}{2N\ell}
   \frac{1}{Z^\ell(\bX^\star)} \prodmu \Big[Z(\mue|\bA^\star_{\ell,\mu})^{n_\mu}\overline{Z(\mue|\bA^\star_{\ell,\mu})}^{m_\mu}\Big] \Big\}.
   \label{eq:generator_before_replica_limits}
    \end{eqnarray}

\subsection{Imaginary replica limits}

At this stage we can safely take the three limits defined in  (\ref{eq:imaginary_limits}), where first for each discretized eigenvalue $\mu$ of the adjacency matrix  the replica dimensions $n_\mu$ and $m_\mu$ take specific imaginary values, followed by the limit $\Delta\to 0$ that converts discretized eigenvalues of adjacency matrices into  continuous ones. The objects in (\ref{eq:generator_before_replica_limits}) affected by these limits are all of the following form, with $\hat{\varrho}^\prime(\mu)=\rmd\hat{\varrho}(\mu)/\rmd\mu$:
\begin{eqnarray}
\hspace*{-15mm}
 \lim  ~\prod_\mu f(\mu)^{n_\mu} \overline{f(\mu)}^{m_\mu} &=& 
    \lim_{\Delta\to 0}~\lim_{n_\mu \to \rmi \frac{\Delta}{\pi}\hrho^\prime(\mu)}~\lim_{m_\mu \to -n_\mu} \rme^{\sum_\mu \big[n_\mu \log f(\mu)+ m_\mu \log \overline{f(\mu)} \big]}
    \nonumber
    \\
    &=&
   \rme^{\frac{2}{\pi}\int\!\rmd\mu~ \hrho(\mu) \dmu \Imag \log f(\mu) }. 
  \end{eqnarray}
  In particular, application to $f(\mu)=[2\pi/\rmi x^\star(\mu)]^{\frac{1}{2}}$ and to $f(\mu)=Z(\mue|\bA^\star_{\ell,\mu})$ gives
  \begin{eqnarray}
  \lim \log Z(\bX^*) &=& -\frac{1}{\pi}\int\!\rmd\mu~ \hrho(\mu) \dmu \Imag \log x^\star(\mu) 
 \end{eqnarray}
 and
 \begin{eqnarray}
 \hspace*{-10mm}
  \lim \prodmu \Big[Z(\mue|\bA^\star_{\ell,\mu})^{n_\mu}\overline{Z(\mue|\bA^\star_{\ell,\mu})}^{m_\mu}\Big]
  &=&   \rme^{\frac{2}{\pi}\int\!\rmd\mu~ \hrho(\mu) \dmu \Imag \log Z(\mue|\bA^\star_{\ell,\mu}) }. 
\end{eqnarray}
This gives us
  \begin{eqnarray}
  \hspace*{-10mm}
    \phi[\hrho] &=& \int\!\rmd\mu~ \hrho(\mu)\Big\{  \frac{1}{2\pi} \dmu~{\rm Im}\Big[
    (q\!-\!2)
    \log \Big(x^\star(\mu)\!-\!\frac{1}{x^\star(\mu)}\Big)
-q  \log x^\star(\mu)\Big]\Big\}
        \nonumber
    \\
     \hspace*{-10mm}
    &&+\frac{1}{N}\sum_{\ell=3}^\infty \frac{(q\!-\!1)^{\ell}}{2\ell}~
    \rme^{\frac{1}{\pi}\int\!\rmd\mu~ \hrho(\mu) \dmu \Imag \big[
    \ell \log x^\star(\mu) 
    +2 \log Z(\mue|\bA^\star_{\ell,\mu})\big] }.
    \label{eqLgenerator_before_McKay}
    \end{eqnarray}
    Since $\delta\phi[\hat{\varrho}]/\delta \hat{\varrho}(\mu)= \bra \varrho(\mu|\bA)\ket$, 
    the first line of  (\ref{eqLgenerator_before_McKay}) is the generator of the asymptotic spectrum  in the limit $N\to\infty$, whereas the second line will give us the ${\mathcal O}(N^{-1})$ finite size corrections to the spectrum. 
In \ref{app:MacKay} we show that, upon taking the  limit $\varepsilon\to 0$,  the factor inside the curly brackets in the first line indeed works out to be exactly the Kesten-McKay law (KM) \cite{kesten1959symmetric,McKay1981expected} for random regular graphs:
\begin{eqnarray}
\label{eq:KMlaw}
    \varrho_0 (\mu) &=& \frac{q}{2\pi}\frac{\sqrt{4(q\!-\!1) - \mu^2}}{q^2 - \mu^2}~ \theta\big[2\sqrt{q\!-\!1}-\abs{\mu}\big].
\end{eqnarray}
This shows that, for regular graphs, the deformation of the measure  in the ensemble (\ref{eq:ensemble}) does not alter the resulting spectrum in leading order, but in sub-leading order  ${\mathcal O}(N^{-1})$.  In regular graphs, the Lagrange parameter $\hat{\varrho}(\mu)$ apparently needs to be rescaled further with $N$ to induce a spectrum that in leading order differs from (\ref{eq:KMlaw}), similar to what was found in \cite{lopez2018exactly}.

Having simplified the first line of  (\ref{eqLgenerator_before_McKay}) to $\int\!\rmd\mu~\hat{\varrho}(\mu)\varrho_0(\mu)$,  we now work out further the exponent in the second line of  (\ref{eqLgenerator_before_McKay}). First, in \ref{app:MacKay} we show that 
\begin{eqnarray}
h(\mu)&=&
-\frac{1}{\pi} \dmu \Imag\log \xmu  = \frac{1}{\pi}\frac{\theta\p{2\sqrt{q\!-\!1}-\abs{\mu}}}{\sqrt{4(q\!-\!1) - \mu^2 }}.
\label{eq:hdistribution}
\end{eqnarray}
We can evaluate the second term in the exponent using the eigenvalues $\lambda_k=2\cos(2\pi k/\ell)$ of the adjacency matrix $A_{ij}=\delta_{i,j+1}+\delta_{i,j-1}~({\rm mod}~\ell)$ of a length-$\ell$ loop, and the identity $\rmd x^\star(\mu)/\rmd\mu=-\rmi x^\star(\mu)/\sqrt{4(q\!-\!1)\!-\!\mu^2}$:
\begin{eqnarray}
\hspace*{0mm}
g_\ell(\mu)&=& \lim_{\varepsilon\downarrow 0}\frac{2}{\ell\pi} \Imag\Big\{\dmu 
 \log Z(\mue|\bA^\star_{\ell,\mu})
 \Big\}
 \nonumber
 \\
 \hspace*{0mm}
 &=& - \frac{1}{\ell\pi} \sum_{k=1}^\ell \Imag\Big\{\dmu 
 \log \Big[\rmi \Big(\lambda_k\!+\!\frac{2\!-\!q}{x^\star(\mu)}\!-\!\mu\Big)\Big]
 \Big\}
 \nonumber
 \\
 \hspace*{0mm}
 &=& \frac{1}{\ell\pi} \sum_{k=1}^\ell \Imag\Big\{
\frac{1\!-\! (2\!-\! q)\frac{\rmd}{\rmd\mu}(x^\star(\mu))^{-1}}{  \lambda_k\!+\!\frac{2-q}{x^\star(\mu)}\!-\!\mu}
 \Big\}
  \nonumber
 \\
 \hspace*{0mm}
 &=& \frac{1}{\ell\pi}  \Imag\Big\{\sum_{k=1}^\ell
\frac{1\!+\! \rmi \frac{q-2}{x^\star(\mu)}[4(q\!-\!1)\!-\!\mu^2]^{-\frac{1}{2}}}{ 2\cos(2\pi k/\ell)\!-\!\frac{q-2}{x^\star(\mu)}\!-\!\mu}
 \Big\}.
 \label{eq:glfunction}
\end{eqnarray}
With the above simplifications we can write both the leading two orders in $N$ of the generating function $\phi[\hat{\varrho}]$ and of the resulting average spectrum $\varrho(\mu)=\delta\phi[\hat{\varrho}]/\delta\hat{\varrho}(\mu)$ for our ensemble (\ref{eq:spectral_ensemble}), for Lagrange parameters $\hat{\varrho}(\mu)={\mathcal O}(1)$,  in the following transparent form, which represents 
one of the main results of this paper:
  \begin{eqnarray}
  \hspace*{-15mm}
    \phi[\hrho] &=& \int\!\rmd\mu~ \hrho(\mu)\varrho_0(\mu)      +\frac{1}{N}\sum_{\ell=3}^\infty \frac{(q\!-\!1)^{\ell}}{2\ell}~
    \rme^{\ell\int\!\rmd\mu~ \hrho(\mu)[g_\ell(\mu)-h(\mu)]}    + o(\frac{1}{N}),
    \label{eq:finalPhi}
    \\
     \hspace*{-15mm}
    \varrho(\mu)&=& \varrho_0(\mu) +\frac{1}{2N}\sum_{\ell=3}^\infty (q\!-\!1)^{\ell}
    \rme^{\ell\int\!\rmd\mu~ \hrho(\mu)[g_\ell(\mu)-h(\mu)]}    [g_\ell(\mu)\!-\!h(\mu)]+ o(\frac{1}{N}).
    \label{eq:correctionsSpectrum}
    \end{eqnarray}
The sum over $\ell$ need not always be convergent, but this has been shown not to necessarily pose problems \cite{ferrari2013finite,lucibello2014finite,metz2014finite}.
The coefficient $(q\!-\!1)^\ell /2\ell$ in $\phi[\hat{\varrho}]$ is exactly the asymptotic expected number of loops of \emph{finite} length $\ell$ inside a random regular graph \cite{wormald1981asymptotic}. It is amazing that this can be recovered with a replica calculation, and gives an intuitive interpretation of the series in (\ref{eq:correctionsSpectrum}): $\ell [g_\ell(\mu) \!- \!h(\mu)]$ is the correction to the Kersten-McKay spectrum formula due to the appearance of a loop of length $\ell$, as discussed in \cite{lucibello2014finite,metz2014finite}. In our present ensemble (\ref{eq:spectral_ensemble}), the number of loops of a given length are given by the usual number found in random regular graphs, multiplied by a factor that depends on the spectral Lagrange parameter $\hrho(\mu)$. Given that the effect of the loops is additive in (\ref{eq:correctionsSpectrum}), we must expect that these spectral corrections come from isolated and well separated loops in the graph.  We wish to point out that, while a cavity approach could account for the presence of loops, it would not be able to provide information on their average number in an ensemble such as (\ref{eq:spectral_ensemble}). The imaginary replica approach presented here, in contrast, has simultaneously provided for the ensemble (\ref{eq:spectral_ensemble}) both the spectrum formula  \emph{and} the expected number of loops.

\subsection{Remaining integrals over eigenvalues}

In our present theory we have an as yet arbitrary functional Lagrange parameter $\hat{\varrho}(\mu)$ which controls the dependence of the graph probabilities on their expected spectra. In  (\ref{eq:correctionsSpectrum}) we still have  integrals over $\hat{\varrho}(\mu)$, of the  form:
\begin{eqnarray}
{\cal J}_\ell[\hat{\varrho}]&=&
  \int\!\rmd\mu~\hrho(\mu) [g_\ell(\mu)\! -\! h(\mu)].
\end{eqnarray}
While expressions (\ref{eq:hdistribution},\ref{eq:glfunction}) for $h(\mu)$ and $g_\ell(\mu)$ will turn out useful in establishing links with previous research in a subsequent section, here we will continue the further evaluation of ${\cal J}_\ell[\hat{\varrho}]$ using the earlier forms
\begin{eqnarray}
h(\mu)&=&-\frac{1}{\pi} \dmu \Imag\log \xmu,
\label{eq:hmu_general}
\\
g_\ell(\mu)&=&- \frac{1}{\pi} \dmu \frac{1}{\ell}\sum_{k=1}^\ell \Imag
 \log \Big[\rmi \Big(\cos(2\pi k/\ell)\!+\!\frac{2\!-\!q}{x^\star(\mu)}\!-\!\mu\Big)\Big].
 \label{eq:glmu_general}
\end{eqnarray}
These give
\begin{eqnarray}
  \hspace*{-15mm} 
{\cal J}_\ell[\hat{\varrho}]
&=&-\Imag
 \int\!\rmd\mu~\hrho(\mu) \frac{1}{\ell \pi}\sum_{k=1}^\ell\dmu  \Big[
 \log \Big(\cos(2\pi k/\ell)\!+\!\frac{2\!-\!q}{x^\star(\mu)}\!-\!\mu\Big)
 -
\log \xmu
 \Big]
\nonumber
\\
\hspace*{-15mm}
&=& \Imag
 \int\!\rmd\mu~\hrho(\mu) \frac{2}{\ell\pi}\sum_{k=1}^\ell \dmu  
\log \Big[\frac{1}{x^\star(\mu)}\Big(\cos(2\pi k/\ell)\!+\!\frac{2\!-\!q}{x^\star(\mu)}\!-\!\mu\Big)\Big]^{-\frac{1}{2}}.
  \end{eqnarray}
We now use 
$\xmu + {\xmu}^{-1} = -\mu - (q\!-\!2)/\xmu$, which follows directly from (\ref{eq:xstar}):
\begin{eqnarray}
  \hspace*{-5mm} 
{\cal J}_\ell[\hat{\varrho}]
&=& \Imag
 \int\!\rmd\mu~\hrho(\mu) \frac{2}{\ell\pi}\sum_{k=1}^\ell \dmu  
\log \Big(1\!+\!\frac{\cos(2\pi k/\ell)}{x^\star(\mu)}\!+\!\frac{1}{(\xmu)^{2}}\Big)^{-\frac{1}{2}}.
\label{eq:integral_before_polynomials}
\end{eqnarray}
In this expression we recognize the generating function of the Chebyshev polynomials $T_n(t)$ \cite{gradshteyn2014table}. These are defined for  $t\in[-1,1]$, and can be written in explicit form as 
 \begin{eqnarray}
    T_n(t)  &=&   \real\p{t \!+\! \rmi\sqrt{1\!-\!t^2}}^n. 
\end{eqnarray}
    They obey the orthogonality relation
\begin{eqnarray}
    \frac{2}{\pi} \int_{-1}^1\frac{\rmd t}{\sqrt{1-t^2}}~T_n(t)T_m(t) &=& \delta_{n m} (1 + \delta_{m 0}) 
    \label{eq:othogonality}
\end{eqnarray}
 as well as
 \begin{eqnarray}
    T_n(\cos(\theta)) = \cos(n\theta),~~~~~~
    T_n(-t) = (-1)^n T_n(t).
\end{eqnarray}
The first five
Chebyshev polynomials are \cite{gradshteyn2014table}:
\begin{eqnarray}
\begin{array}{lll}
T_0(t)=1, &~~ T_1(t)=t, &~~ T_{2}(t)=2t^2-1,
\\[1mm]
 T_{3}(t)=4t^3-3t, &~~
T_{4}(t)=8t^4-8t^2+1.
\end{array}
\end{eqnarray}
 For the evaluation of (\ref{eq:integral_before_polynomials}), in particular, we may apply the generating function identity
 \begin{eqnarray}
    \sum_{n=1}^\infty T_n(t)\frac{x^n}{n}  &=& \log\p{1 - 2 t x  + x^2}^{-\frac{1}{2}} ~~~~~\textrm{for }\abs{x}<1
    \end{eqnarray}
    to the choices $x=1/x^\star(\mu)$ and $t=-\cos(2\pi k/\ell)$, in order to obtain
\begin{eqnarray}
  \hspace*{-5mm} 
{\cal J}_\ell[\hat{\varrho}]
&=& \Imag
 \int\!\rmd\mu~\hrho(\mu) \frac{2}{\ell\pi}\sum_{k=1}^\ell 
  \sum_{n=1}^\infty \frac{(-1)^n}{n}T_n(\cos(2\pi k/\ell)) \dmu   (x^\star(\mu))^{-n}
  \nonumber
  \\
  &=& \Imag
 \int\!\rmd\mu~\hrho(\mu) \frac{2}{\ell\pi}\sum_{k=1}^\ell 
  \sum_{n=1}^\infty (-1)^{n+1} \frac{T_n(\cos(2\pi k/\ell)) }{  (x^\star(\mu))^{n+1}}\frac{\rmd}{\rmd\mu}x^\star(\mu).
  \end{eqnarray}
We next use $\rmd x^\star(\mu)/\rmd\mu=-\rmi x^\star(\mu)/\sqrt{4(q\!-\!1)\!-\!\mu^2}$ and the short-hands 
\begin{eqnarray}
    d_{n,\ell} &=&\frac{1}{\ell}\sum_{k=1}^\ell T_n(\cos(2\pi k/\ell)) \nonumber
    \\&=& \frac{1}{\ell}\sum_{k=1}^\ell \cos(2\pi nk/\ell)
     = \sum_{p\in\mathbb{Z}} \delta_{n,p\ell}.
\end{eqnarray}
This results in
\begin{eqnarray}
  \hspace*{-5mm} 
{\cal J}_\ell[\hat{\varrho}]
&=& \Imag
 \int\!\rmd\mu~\hrho(\mu) \frac{2}{\pi}
  \sum_{n=1}^\infty (-1)^{n} \frac{d_{n,\ell}}{  (x^\star(\mu))^{n}}
  \frac{\rmi}{\sqrt{4(q\!-\!1)\!-\!\mu^2}}
  \nonumber
  \\
  &=&   \sum_{n=1}^\infty (-1)^{n} d_{n,\ell} ~
 \frac{2}{\pi}\int\!\rmd\mu~\hrho(\mu) {\rm Re}\Big[
  \frac{1}{  (x^\star(\mu))^{n}}
  \frac{1}{\sqrt{4(q\!-\!1)\!-\!\mu^2}}\Big]
   \nonumber
  \\
  &=&   \sum_{n=1}^\infty (-1)^{n} d_{n,\ell} ~
 \frac{2}{\pi}\int\!\rmd\mu~\frac{\hrho(\mu) }{|x^\star(\mu)|^{2n}}{\rm Re}\Big[
  \frac{\overline{x^\star(\mu)}^{n}}{\sqrt{4(q\!-\!1)\!-\!\mu^2}}\Big].
  \end{eqnarray}
 If $\mu^2<4(q\!-\!1)$ one has $|x^\star(\mu)|^2=q-1$ and  $\sqrt{4(q\!-\!1)\!-\!\mu^2}\in[0,\infty)$. For $\mu^2>4(q\!-\!1)$, on the other hand, we have $x^\star(\mu)\in{\rm I\!R}$ and $\sqrt{4(q\!-\!1)\!-\!\mu^2}$ is purely imaginary. 
 We also note that for $\mu^2<4(q\!-\!1)$ we may write 
 \begin{eqnarray}
 {\rm Re}\Big(
 \overline{x^\star(2t\sqrt{q\!-\!1})}^{n}\Big)&=& (q\!-\!1)^{n/2}~{\rm Re}\Big(\!-\!t\!+\!\rmi\sqrt{1\!-\!t^2}\Big)^n
 \nonumber
 \\
 &=& (-1)^n (q\!-\!1)^{n/2}~T_n(t).
  \end{eqnarray}
 In combination these properties allow us to simplify ${\cal I}_\ell[\hat{\varrho}]$ to
  \begin{eqnarray}
{\cal J}_\ell[\hat{\varrho}]
&=& \sum_{n=1}^\infty \frac{(-1)^{n} d_{n,\ell}}{(q\!-\!1)^n} ~
 \frac{2}{\pi}\int_{-2\sqrt{q-1}}^{2\sqrt{q-1}}\!\frac{\rmd\mu~\hrho(\mu) {\rm Re}\big[
 \overline{x^\star(\mu)}^{n}\big]}{\sqrt{4(q\!-\!1)\!-\!\mu^2}}
 \nonumber
 \\
 &=& \sum_{n=1}^\infty \frac{d_{n,\ell}}{(q\!-\!1)^{n/2}} ~
 \frac{2}{\pi}\int_{-1}^{1}\!\frac{\rmd t}{\sqrt{1-t^2}}~\hrho(2t\sqrt{q\!-\!1})
T_n(t)
\nonumber
 \\
 &=&  \sum_{p=1}^\infty  \frac{1}{(q\!-\!1)^{p\ell/2}} ~
 \frac{2}{\pi}\int_{-1}^{1}\!\frac{\rmd t}{\sqrt{1-t^2}}~\hrho(2t\sqrt{q\!-\!1})
T_{p\ell}(t).
\label{eq:integral_arbitrary_rhohat}
  \end{eqnarray}
  The above result shows that the Chebyshev polynomials form the natural basis in terms of which to express the  functional Lagrange parameter $\hat{\varrho}(\mu)$, and  is inserted into our spectrum formula
    (\ref{eq:correctionsSpectrum}) to give
    \begin{eqnarray}
     \varrho(\mu)&=& \varrho_0(\mu) +\frac{1}{2N}\sum_{\ell=3}^\infty (q\!-\!1)^{\ell}
    \rme^{\ell {\cal J}_\ell[\hat{\varrho}]}    [g_\ell(\mu)\!-\!h(\mu)]+ o(\frac{1}{N}).
    \label{eq:spectrum_with_J}
    \end{eqnarray} 
It is instructive to work out ${\cal J}_\ell[\hat{\varrho}]$ for $\ell\geq 3$ and some simple choices of $\hat{\varrho}(\mu)$:
\begin{itemize}
    \item $\hrho(\mu) = \mu^3$:
    \\[2mm]
    This choice corresponds to random regular graphs in which the number of triangles is controlled. 
    We use $t^3=\frac{1}{4}T_3(t)+\frac{3}{4}T_1(t)$ and the orthogonality relation   (\ref{eq:othogonality}):
     \begin{eqnarray}
     \hspace*{-15mm}
{\cal J}_\ell[\hat{\varrho}]
&=&  \sum_{p=1}^\infty  \frac{2}{(q\!-\!1)^{(p\ell-3)/2}} ~
 \frac{2}{\pi}\int_{-1}^{1}\!\frac{\rmd t}{\sqrt{1-t^2}}\big[T_3(t)\!+\!3T_1(t)
\big] 
T_{p\ell}(t)
\nonumber
\\
 \hspace*{-15mm}
&=&  2\sum_{p=1}^\infty  \frac{\delta_{3,p\ell}+3\delta_{1,p\ell}}{(q\!-\!1)^{(p\ell-3)/2}}
=  2\sum_{p=1}^\infty  \delta_{3,p\ell} =  2\delta_{\ell 3}.
\label{eq:Jintegral_triangles}
  \end{eqnarray}
         
    \item $\hrho(\mu) = \mu^4$:
    \\[2mm]
     This choice corresponds to random regular graphs in which the number of squares is controlled. 
   We use $t^4=\frac{1}{8}T_4(t)+\frac{1}{4}T_2(t)+\frac{3}{8}T_0(t)$ and the orthogonality  (\ref{eq:othogonality}):
         \begin{eqnarray}
          \hspace*{-15mm}
{\cal J}_\ell[\hat{\varrho}]
&=&   \sum_{p=1}^\infty  \frac{2}{(q\!-\!1)^{(p\ell-4)/2}} ~
 \frac{2}{\pi}\int_{-1}^{1}\!\frac{\rmd t}{\sqrt{1-t^2}}
 \big[T_4(t)+2T_2(t)+3T_0(t)\big]
T_{p\ell}(t)
\nonumber
\\
 \hspace*{-15mm}&=&  2\sum_{p=1}^\infty  \frac{\delta_{4,p\ell}+2\delta_{2,p\ell}+3\delta_{0,p\ell}}{(q\!-\!1)^{(p\ell-4)/2}} 
 = 2\sum_{p=1}^\infty \delta_{4,p\ell}=2\delta_{\ell 4}.
 \label{eq:Jintegral_squares}
  \end{eqnarray}
\end{itemize}

\section{Applications of the general theory}

\subsection{Recovering previous results as a test}

Upon making the trivial choice  $\hrho(\mu) = 0$ we return to the conventional ensembles with uniform probabilities, and our equations (\ref{eq:finalPhi},\ref{eq:correctionsSpectrum}) recover the natural spectrum fluctuations of random regular graphs, as previously studied in detail in \cite{johnson2011exchangeable} and with the traditional replica method (where $n\to 0$) in \cite{metz2014finite}: 
\begin{eqnarray}
    \varrho(\mu) &= & \varrho_0(\mu) + N^{-1}\varrho_1(\mu)+o(N^{-1})
\end{eqnarray}
with 
\begin{eqnarray}
\varrho_1(\mu)&=&
\sum_{\ell = 3}^\infty \frac{(q\!-\!1)^\ell}{2} [g_\ell(\mu) \!-\! h(\mu)].
\label{eq:rho1}
\end{eqnarray}
This series was summed in \cite{metz2014finite},  and we can connect the result of the summation, in the notation of \cite{metz2014finite}, directly to the theory developed in the present paper as follows:
\begin{eqnarray}
\label{eq:resum}
    \varrho_1(\mu) &= & h(\mu) ~\real \s{\frac{(q\!-\!1) g_c(\mu)}{1\!-\!(q\!-\!1)g_c(\mu)}  + \frac{(q\!-\!1)g_c^2(\mu)}{1\!-\! (q\!-\!1)g_c^2(\mu)}}\nonumber\\
    && + h(\mu)~\real\s{ \sum_{\ell=3}^\infty (q\!-\!1)^\ell \frac{g_c^{3\ell}(\mu)}{ 1\!- \!g_c^\ell(\mu)} \!-\! K(g_c(\mu))},
    \label{eq:resummation}
    \end{eqnarray}
    in which now
    \begin{eqnarray}
    g_c(\mu) &=& -1/\xmu,
    \\
    K(g) &=& (q\!-\!1)g + q(q\!-\!1)g^2 + (q\!-\!1)^2 g^4.
\end{eqnarray}
Here $h(\mu)$ and $x^\star(\mu)$ are given in (\ref{eq:hdistribution}) and (\ref{eq:xstarmu}), respectively.
\vsp

As a second test we can make the special choices $q=2$ and  $\hat{\varrho}(\mu)=\alpha \mu^3$,  resulting in the ensemble that was studied in \cite{lopez2018exactly} via direct combinatorics, i.e.  without the replica method. This particular model represents  the simplest solvable non-uniform random graph ensemble with tuneability of the frequency of short loops. First, by setting $q=2$ our general results 
(\ref{eq:hdistribution},\ref{eq:glfunction},\ref{eq:correctionsSpectrum}) 
simplify greatly. We now find that 
\begin{eqnarray}
\hspace*{-15mm}
    \varrho_0 (\mu) &=& \frac{1}{\pi}\frac{\theta(2-\abs{\mu})}{\sqrt{4\! -\! \mu^2}},~~~~~~ h(\mu) =\varrho_0 (\mu) 
    \\
    \hspace*{-15mm}
    g_\ell(\mu)  &=& \lim_{\varepsilon\downarrow 0}\frac{1}{\ell\pi}  \Imag\Big\{\sum_{k=1}^\ell
\frac{1}{ 2\cos(2\pi k/\ell)\!-\!\mu\!-\!\rmi\varepsilon}
 \Big\}
= \frac{1}{\ell} \sum_{k=1}^\ell
\delta \Big[\mu\!-\!2\cos(\frac{2\pi k}{\ell})\Big]
\end{eqnarray}
Upon inserting also $\hat{\varrho}(\mu)=\alpha \mu^3$ into  (\ref{eq:correctionsSpectrum}) 
 we need the values of
\begin{eqnarray}
\int\!\rmd\mu~\mu^3  \varrho_0 (\mu) &=& 0
\\
\int\!\rmd\mu~\mu^3  g_\ell (\mu) &=& \frac{8}{\ell} \sum_{k=1}^\ell 
\cos^3(\frac{2\pi k}{\ell})=2\delta_{\ell 3}+8 \delta_{\ell 1}
 \end{eqnarray}
 With this the ensemble spectrum becomes
\begin{eqnarray}
\hspace*{-22mm}
 \varrho(\mu)&=& \varrho_0(\mu) +\frac{1}{2N}\sum_{\ell=3}^\infty 
    \rme^{6\alpha \delta_{\ell 3}}    \Big\{
     \frac{1}{\ell} \sum_{k=1}^\ell
\delta \Big[\mu\!-\!2\cos(\frac{2\pi k}{\ell})\Big]
    - \frac{1}{\pi}\frac{\theta(2\!-\!\abs{\mu})}{\sqrt{4\! -\! \mu^2}}
    \Big\}+ o(\frac{1}{N})
      \end{eqnarray}
and for the triangle density $ m (\alpha) = \int\!\rmd\mu~\varrho(\mu) \mu^3 $ we obtain
\begin{eqnarray}
    m (\alpha) =  N^{-1}\rme^{6\alpha}
\end{eqnarray}
These results are indeed identical to those derived combinatorially in \cite{lopez2018exactly}.

\subsection{Triangularly constrained regular graph ensemble with arbitrary degree}

\begin{figure}[t]

\begin{picture}(372,140)
    \put(20,5){\includegraphics[width=0.5\textwidth]{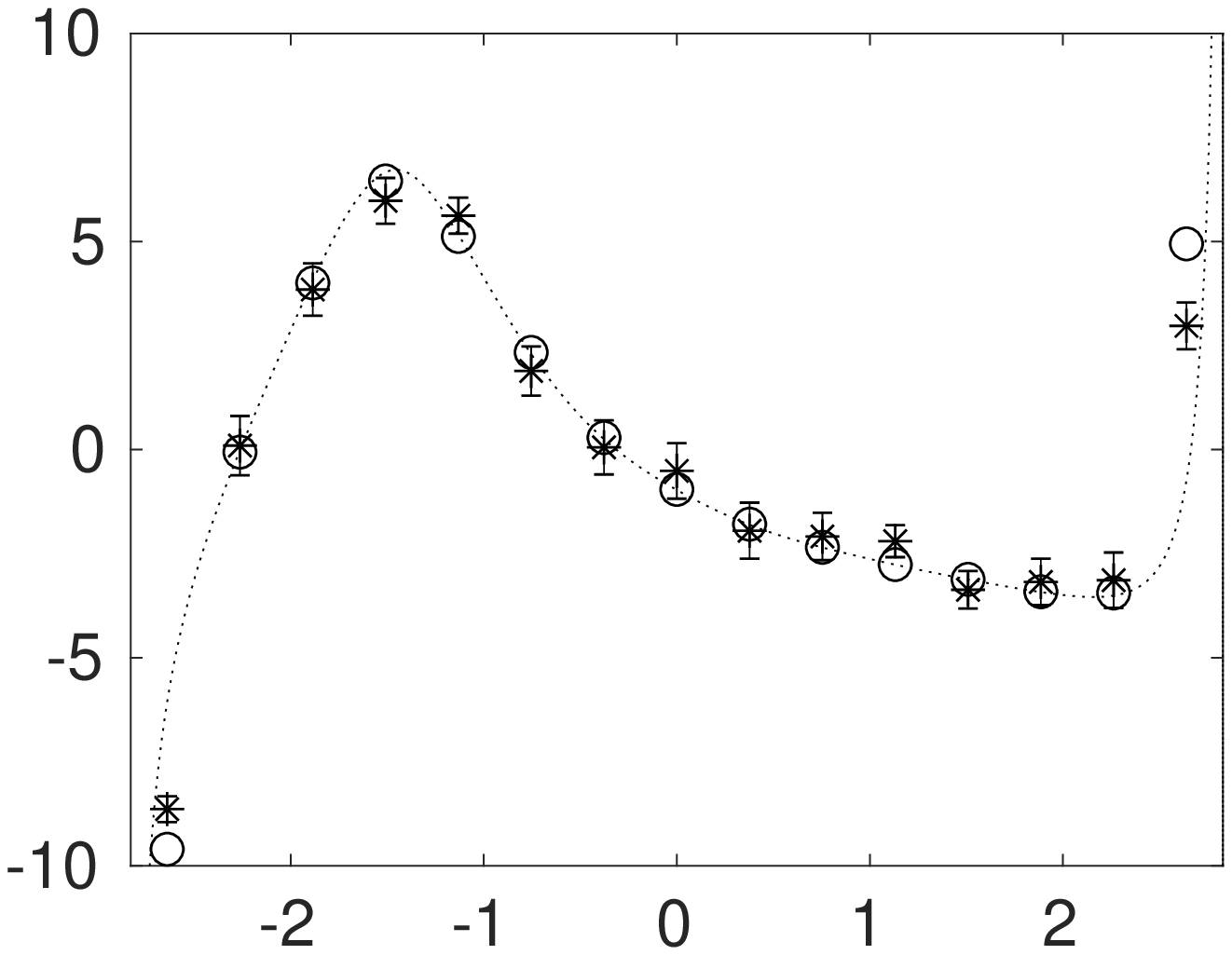}}
    \put(200,5){\includegraphics[width=0.5\textwidth]{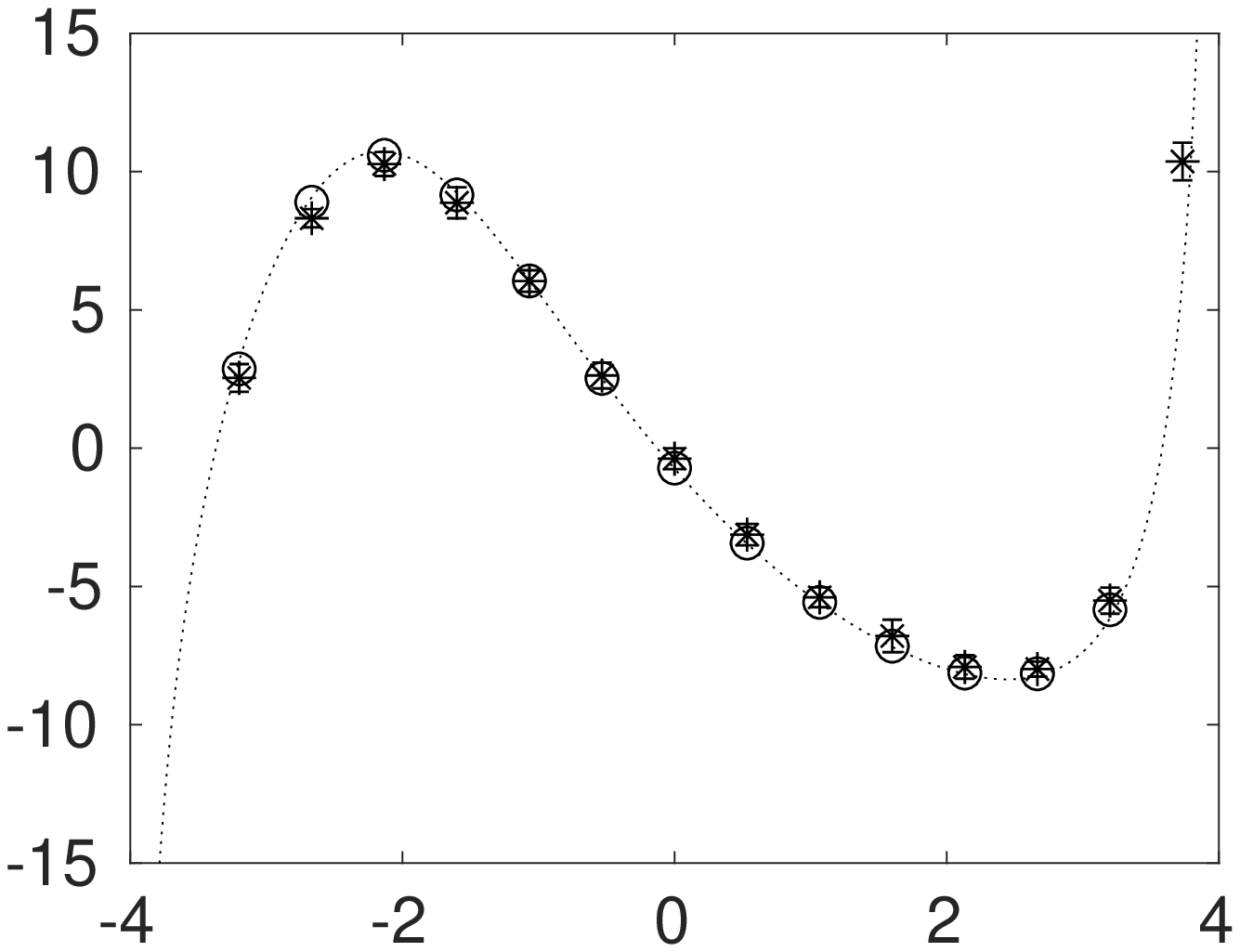}}
    \put(150,120){$q=3$}   \put(330,120){$q=5$}
    \put(113,0){\small $\mu$}
    \put(293,0){\small $\mu$}
    \put(6,75){$\delta \varrho(\mu)$}
\end{picture}

\caption{\label{fig:loopySpectra} Average spectral densities for $q$-regular graphs sampled from (\ref{eq:ensemble}). We show the rescaled finite size deviations from the standard Kersten-McKay formula $\varrho_0(\mu)$, by plotting $\delta\varrho(\mu) = N[\varrho(\mu)-\varrho_0(\mu)]=\varrho_1(\mu)+\tilde{\varrho}_1(\mu)$. Left panel:    $q=3$, $N=1000$ and $\alpha=0.416$, giving average clustering coefficient $\bra C(\bA)\ket=0.016$. Right figure:  $q=5$, $N=2000$ and $\alpha=0.431$, giving average clustering coefficient $\bra C(\bA)\ket= 0.02$.  Each marker shows the average spectral density contribution obtained from 200 histograms of samples of (\ref{eq:ensemble}), generated with an appropriate MCMC process, and error bars indicate $\pm$ one standard deviation. The dotted line shows the theoretical prediction (\ref{eq:triangle_spectrum},\ref{eq:rho1tilde}), and circles show the density prediction computed for exactly the eigenvalue bins that were also used for the histograms of the simulation samples.  } 
\end{figure}

We proceed to apply the general theory developed in the previous section to the graph ensemble (\ref{eq:ensemble}) with controlled numbers of triangles, i.e. with $\hat{\varrho}(\mu)=\alpha\mu^3$, but now for arbitrary values of the degree $q$ where the direct combinatorial approach of \cite{lopez2018exactly} is no longer possible. 
We can start directly by inserting  (\ref{eq:Jintegral_triangles}) into (\ref{eq:spectrum_with_J}), upon adding the control  parameter $\alpha$, giving
  \begin{eqnarray}
  \hspace*{-0mm}
     \varrho(\mu)&=& \varrho_0(\mu) +\frac{1}{2N}\sum_{\ell=3}^\infty (q\!-\!1)^{\ell}
    \rme^{ 6\alpha \delta_{\ell 3}}    [g_\ell(\mu)\!-\!h(\mu)]+ o(\frac{1}{N})
    \nonumber
    \\
     \hspace*{-0mm}
   &=&  \varrho_0(\mu)  +\frac{1}{N}\varrho_1(\mu)+\frac{1}{N}\tilde{\varrho}_1(\mu) + o(\frac{1}{N}),
   \label{eq:triangle_spectrum}
   \\
   \hspace*{-0mm}
   \tilde{\varrho}_1(\mu)&=&
   \frac{1}{2} (q\!-\!1)^{3}
   ( \rme^{ 6\alpha}\!-\!1)    [g_3(\mu)\!-\!h(\mu)].   
   \label{eq:rhotilde}
  \end{eqnarray}
 Here $\varrho_1(\mu)$ is the function (\ref{eq:rho1}) that already appeared in the spectrum of the non-deformed ensembles of 
\cite{metz2014finite}, and for which we can use the resummation   (\ref{eq:resummation}). 
The impact of controlling the graph probabilities (\ref{eq:ensemble})  with a nonzero Lagrange parameter $\hat{\varrho}(\mu)=\alpha\mu^3$ 
is fully concentrated in $\tilde{\varrho}_1(\mu)$.
We next insert our earlier expressions for $g_\ell(\mu)$ and $h(\mu)$ into (\ref{eq:rhotilde}) and simplify the result where possible:
    \begin{eqnarray}
  \hspace*{-20mm}
    \tilde{\varrho}_1(\mu) 
   &=& \frac{(q\!-\!1)^{3}}{2\pi} 
   ( \rme^{ 6\alpha}\!-\!1)~ \theta[2\sqrt{q\!-\!1}\!-\!\abs{\mu}]
   \nonumber
   \\[-1mm]
    \hspace*{-20mm}
   && \hspace*{10mm} \times \Big\{
 \Imag\Big[\frac{1}{3}\sum_{k=1}^3
\frac{1\!+\! \rmi \frac{q-2}{x^\star(\mu)}[4(q\!-\!1)\!-\!\mu^2]^{-\frac{1}{2}}}{ 2\cos(2\pi k/3)\!-\!\frac{q-2}{x^\star(\mu)}\!-\!\mu}
 \Big]
-
 \frac{1}{\sqrt{4(q\!-\!1)\! -\! \mu^2 }}
   \Big\}
   \nonumber
   \\[1mm]
    \hspace*{-20mm}
   &=&  \frac{(q\!-\!1)^{3}}{2\pi} 
   ( \rme^{ 6\alpha}\!-\!1) \frac{\theta[2\sqrt{q\!-\!1}\!-\!\abs{\mu}]}{\sqrt{4(q\!-\!1)\! -\! \mu^2 }}
 \Bigg\{
\frac{q\!-\!2}{3}\Big[\frac{2q+\mu}{q^2\!-\!3(q\!-\!1)\!+\!\mu q\!+\!\mu^2}
\!+\!
\frac{1}{q\!-\!\mu}
\Big]
\!-\!
1
   \Bigg\}.
   \nonumber
   \\
   \hspace*{-20mm}&&
   \label{eq:rho1tilde}
 \end{eqnarray} 
  The results of testing this prediction against numerical simulations are shown in Figure \ref{fig:loopySpectra}, and reveal excellent agreement.
In the simulations 
we sampled numerically from (\ref{eq:ensemble}) with an edge swap based Markov Chain Monte Carlo algorithm (MCMC) with nontrivial acceptance probabilities. Edge swaps are accepted or rejected  depending on  the change in the number of loops and on  the change in the possible  number of possible swaps. This corrects for entropic effects, see e.g.  \cite{sampling} or \cite{annibale2017generating}. Since  we work with a system of finite size $N$, our predictions refer to the \emph{average} eigenvalue density,  not to the density of individual graph instances. The error bars in Figure \ref{fig:loopySpectra} are computed following 10 different initialization seeds of the MCMC algorithm, consisting of distinct  regular graphs sampled uniformly with a configuration model algorithm. Following each initialization, 20 samples were taken, separated in algorrithmic time by $\sim 10^3$ accepted MCMC swaps per link in the graph.

We can also calculate the expected triangle density  $m(\alpha) =\int\!\rmd\mu  ~ \varrho(\mu) \mu^3$ for the ensemble (\ref{eq:ensemble}).  It is easier to do this by integrating over    (\ref{eq:spectrum_with_J}) rather than via (\ref{eq:triangle_spectrum},\ref{eq:rho1tilde}), although both routes give the same result:
\begin{eqnarray}
\hspace*{-10mm}
    m(\alpha) 
    &=&\frac{1}{2N}\sum_{\ell=3}^\infty (q\!-\!1)^{\ell}
    \rme^{\alpha \ell  \int\!\rmd\mu~\mu^3  [g_\ell(\mu)\!-\!h(\mu)]}  \int\!\rmd\mu~\mu^3  [g_\ell(\mu)\!-\!h(\mu)]+ o(\frac{1}{N})
   \nonumber
   \\
   \hspace*{-10mm}
   &=& 
  \frac{1}{2N}\sum_{\ell=3}^\infty (q\!-\!1)^{\ell}
    \rme^{2\delta_{\ell 3}\alpha \ell } 2\delta_{\ell 3}+ o(\frac{1}{N})
    \nonumber
  \\    & =& \frac{1}{N}(q-1)^3 \rme^{6\alpha}+ o(\frac{1}{N}).
  \label{eq:triangledensity}
\end{eqnarray}
This formula gives very accurate results for $\alpha$ values up to a certain point, defined as $\alpha_1(N)$ in the next section. This can be seen very clearly in figures \ref{fig:graphs1} and \ref{fig:triangleDensity}, where we test its predictions for ensembles (\ref{eq:ensemble}) with $q=3$. Since $m(\alpha)$ represents an ensemble average, we compare (\ref{eq:triangledensity}) against the average loop density over multiple graphs drawn from the ensemble, $\hat{m}(\alpha) = M^{-1}\sum_{m=1}^M \Tr(\bA_m^3)$. Figures \ref{fig:graphs1} and \ref{fig:triangleDensity} show averages and standard deviations of the estimator $\hat{m}(\alpha)$, for 50 different small samples sampled during MCMC simulations, separated in algorithmic time by $\sim 10^3 $ swaps per link (in order to ensure independence of the $M$ samples). 

We have developed a theory that quantifies the $\Order{1/N}$ effects on the eigenvalue spectrum of probability deformations in ensembles of the general family (\ref{eq:spectral_ensemble}), in which loops can be induced via the functional Lagrange parameter $\hat{\varrho}(\mu)$, and we applied our results to a specific member  (\ref{eq:ensemble}) of this family. As mentioned before, the theory implicitly assumes that loops inside the graph are far away from each other. As the control parameter $\alpha$ in (\ref{eq:ensemble})  is increased for fixed system size $N$, we must therefore expect the behaviour of the ensemble to start deviating from the predictions (\ref{eq:triangle_spectrum},\ref{eq:rho1tilde},\ref{eq:triangledensity}) as soon as the loops start to interact.  This can indeed be seen in figure \ref{fig:graphs1}.  As $\alpha$ increases the clustering coefficient starts deviating from (\ref{eq:triangledensity}), which is shown as a dashed line. Mathematically, one can explain the deviations from (\ref{eq:triangledensity}) as the emergence of higher order corrections to the saddle point approximation, $\Order{N^{-\gamma}}$ with $\gamma>1$, that were not incorporated into the replica calculation. These would account for the presence of loops that are not isolated from each other. The accuracy of (\ref{eq:triangle_spectrum},\ref{eq:rho1tilde}) and  (\ref{eq:triangledensity})  suggests that calculating higher order corrections in the replica calculation would improve our predictions, but this would require of course a much more complicated calculation. In the next section we will explore what happens as we keep increasing $\alpha$ beyond the validity of our replica calculation.

\subsection{Phases of the ensemble and the shattering transition}

\begin{figure}[t]

\begin{picture}(372,230)
  \put(15,0){\includegraphics[width=\textwidth]{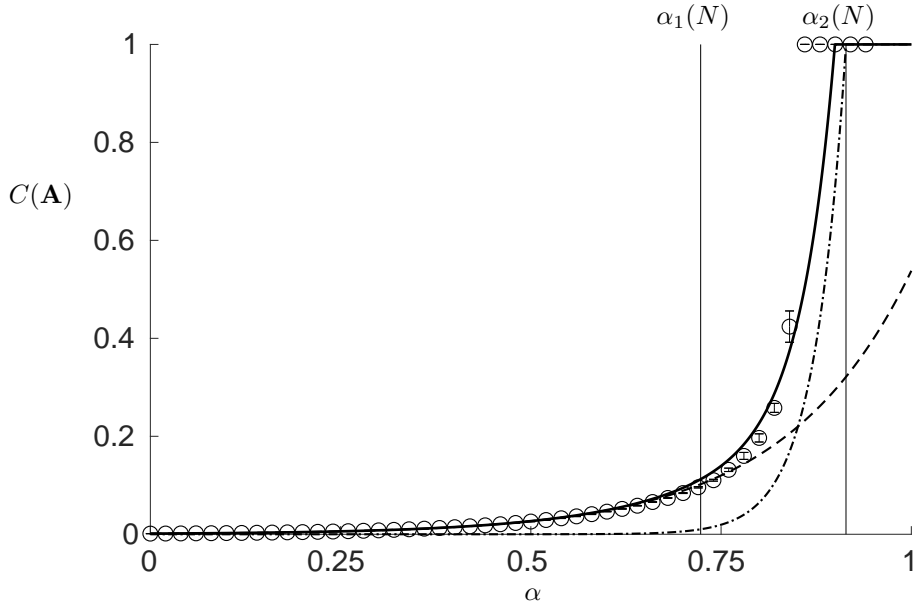}}
  \put(255,218){$\alpha_1(N)$}
  \put(310,218){$\alpha_2(N)$}
  \put(205,0){$\large\alpha$}
  \put(10,150){$C(\bA)$}
\end{picture}

\caption{\label{fig:graphs1} Plot of the clustering coefficient $C(\bA)$ versus $\alpha$. Squares show results from MCMC sampling with $N=1000$ (average plus/minus one standard deviation). Solid line: predicted values computed from  the theory (\ref{eq:triangledensity}), via $\bra C(\bA)\ket=m(\alpha)/q(q\!-\!1)$. We also show separately the two distinct contributions to the theoretical prediction, viz. $\bra C(\bA)\ket_T$ (those from disconnected triangles, dashed line) and $\bra C(\bA)\ket_K$ (those from triangles in cliques, dotted dashed line). Typical graph examples generated within each $\alpha$ regime are shown in Figure \ref{fig:graphs2}.}
\end{figure}

\begin{figure}[t]

\begin{picture}(372,140)
  \put(0,4){\includegraphics[width=0.32\textwidth]{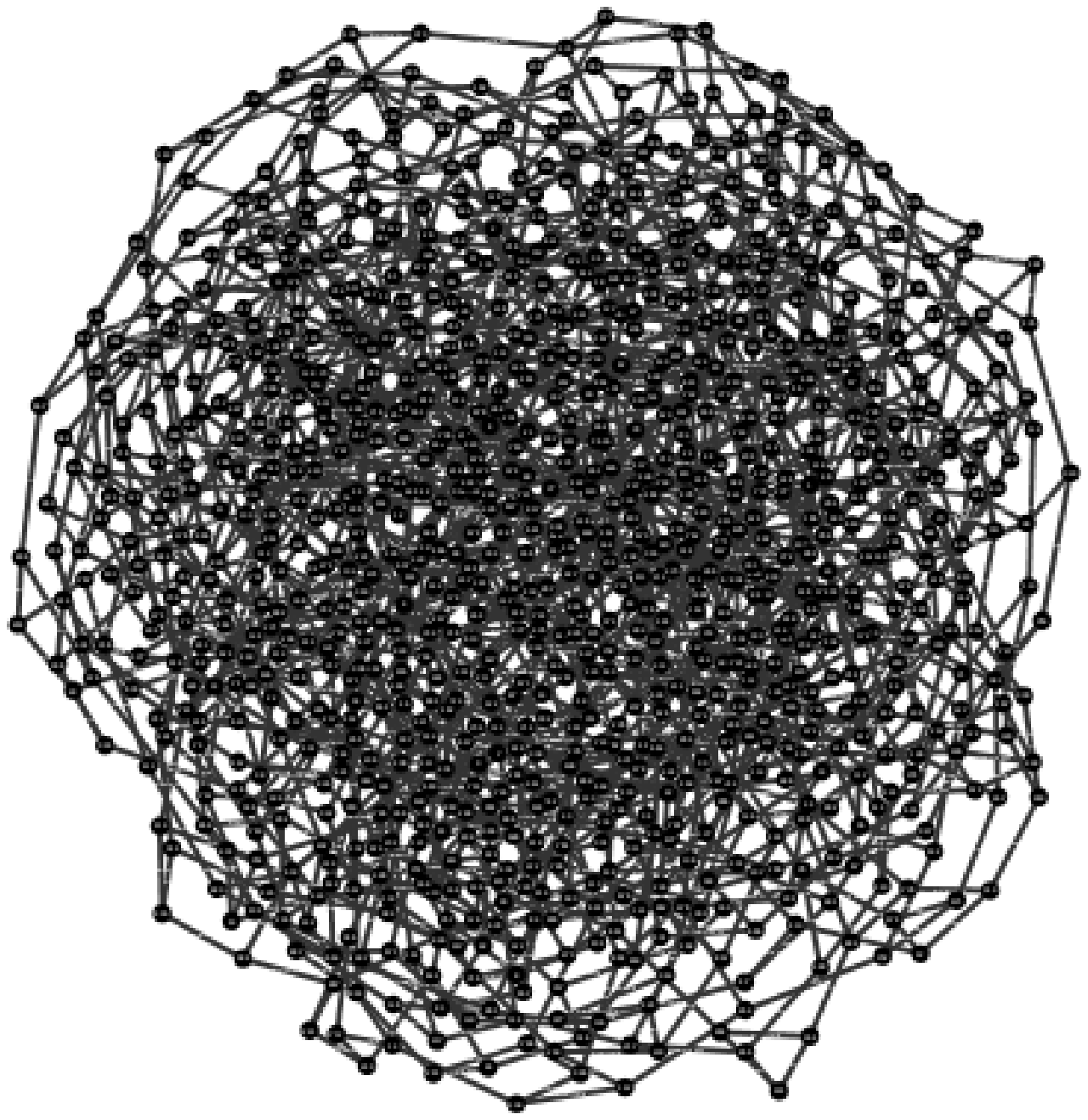}}
  \put(130,4){\includegraphics[width=0.32\textwidth]{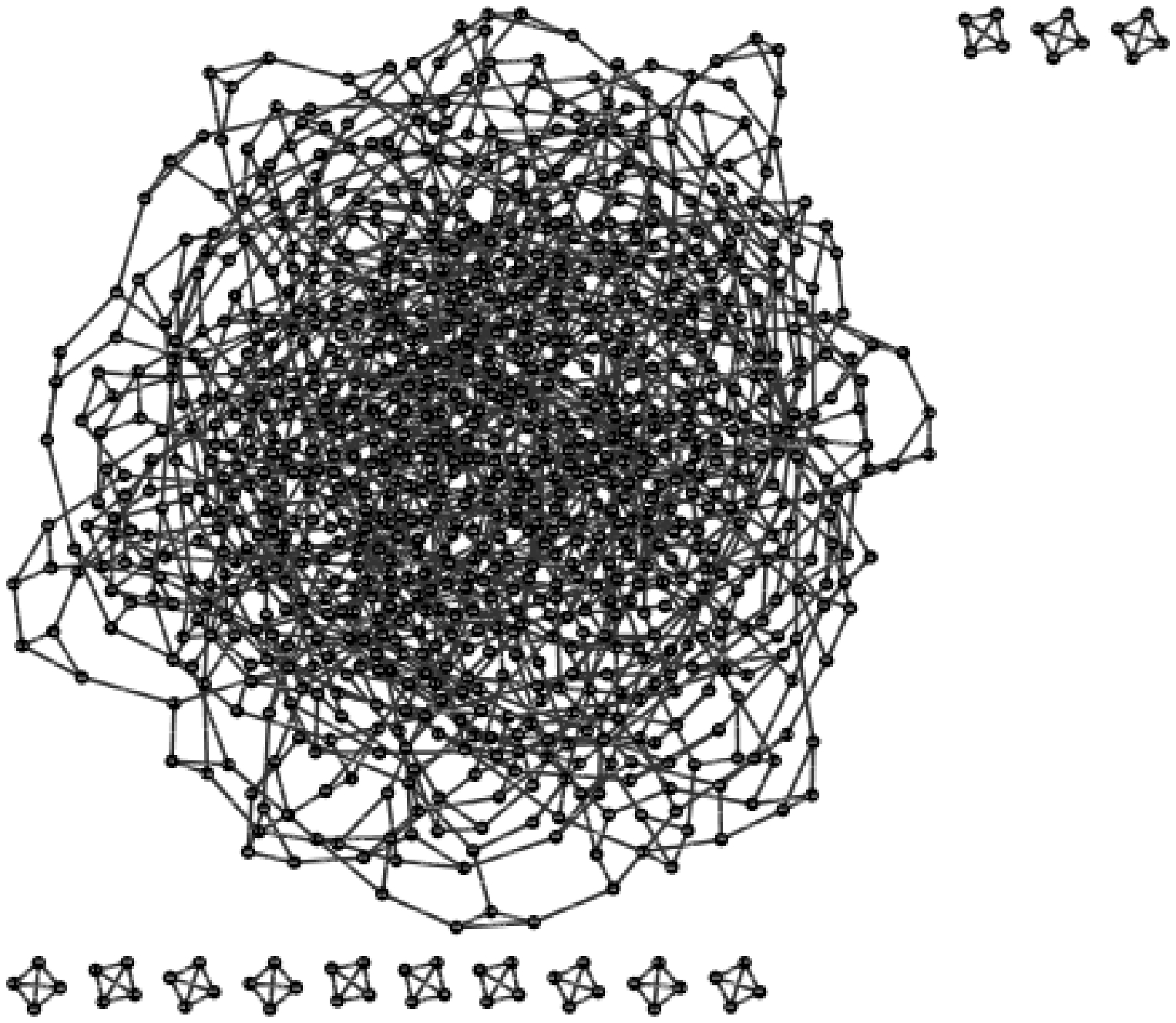}}
  \put(255,4){\includegraphics[width=0.32\textwidth]{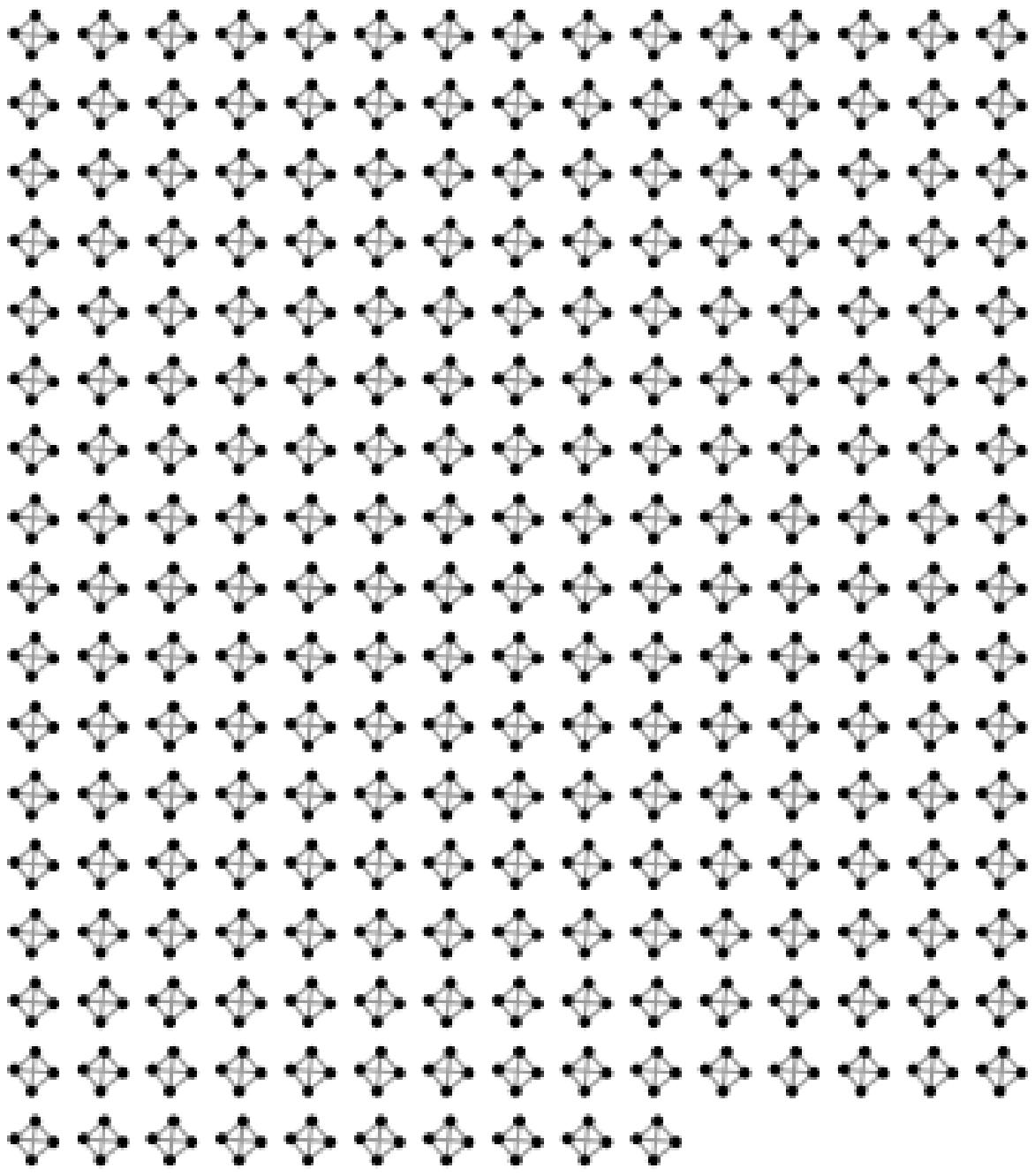}}
\end{picture}

\begin{tikzpicture}
    \draw (0,0) node{$|$} -- (4,0) node{$|$}--(8.75,0)node{$|$} -- (12,0);
    \draw (0,-0.5)node{$0$};
    \draw (4,-0.5)node{$\alpha_1(N)$};
    \draw (8.75,-0.5)node{$\alpha_2(N)$};
    \draw (12.25,0)node{$\infty$};
    \draw (2,0.5)node{\em connected phase};
    \draw (6.5,0.5)node{\em partially connected phase};
    \draw (10.5,0.5)node{\em disconnected phase};
    \draw (4,-1)node{{\em shattering transition}};
\end{tikzpicture}

\caption{\label{fig:graphs2} Three typical 3-regular graphs, of size $N=1000$,  sampled numerically via MCMC from the canonical ensemble (\ref{eq:ensemble}).  The value of the tuning parameter $\alpha$ increases from left to right, and each graph shown is generated from one of the three distinct phases defined in Figure \ref{fig:graphs1}. }
\end{figure}

\begin{figure}[t]
    \centering
    \begin{picture}(350,195)
    \put(30,5){\includegraphics[width=0.7\textwidth]{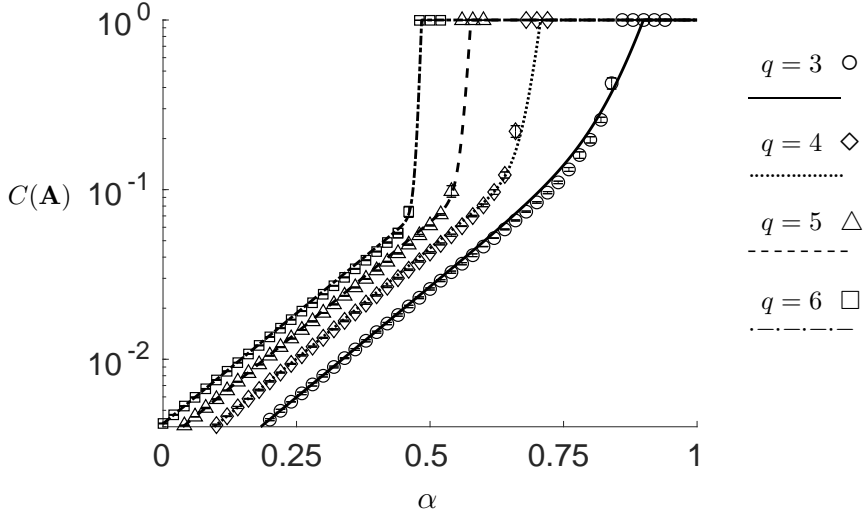}}
    \put(160,-5){\large$\alpha$}
    \put(5,110){$C(\bA)$}
    \put(290,160){$q=3$~ {\Large $\circ$}}
    \put(285,150){\line(1,0){35}}
    \put(290,130){$q=4$~ {\Large$\diamond$}}
    \multiput(285,120)(2,0){18}{.}
    \put(290,100){$q=5$~ $\bigtriangleup$}
    \multiput(285,90)(5,0){8}{-}
    \put(290,70){$q=6$~ $\square$}
    \multiput(285,60)(10,0){4}{$\cdot-$}
    
    \end{picture}\vspace*{2mm}
    
    \caption{\label{fig:triangleDensity} In this figure we show the agreement between the clustering coefficients predicted by (\ref{eq:fullTriangleDensity}),  with lines, and the values measured MCMC simulations (markers, showing average plus/minus one standard deviation) with $N=1000$. Full details on the number of samples generated and their separation in MCMC edge swaps are given in the main text. These results confirm that (\ref{eq:fullTriangleDensity}) captures the essence of the phenomenology of the ensemble.} 
\end{figure}

We will now give a qualitative picture of the behaviour of the ensemble (\ref{eq:ensemble}) for all values of $\alpha \in [0,\infty)$. We will focus on $q\ge 3$, since  the case $q=2$ was already covered in \cite{lopez2018exactly}. In MCMC simulations one observes three distinct regimes, which  are not phases in a rigorous thermodynamic sense, but size dependent ranges of $\alpha$ values that exhibit qualitatively different phenomenology:
\begin{itemize}
    \item {\em Small} $\alpha$: {\em connected phase}
    \\[0.5mm] 
    The triangle promoting probability bias in the ensemble introduces isolated loops embedded in the giant component. Here the analysis of the previous section should apply, as is confirmed in figures \ref{fig:graphs1} and \ref{fig:triangleDensity} for different values of $q$. Indeed one observes only small deviations from (\ref{eq:triangledensity}), as one approaches the next phase.
    \item {\em Intermediate} $\alpha$: {\em partially connected phase}
    \\[0.5mm]
    Edges can now be part of more than one triangle, and the graphs contain an increasing number of cliques of $q+\!1$ nodes, denoted by $K_{q+1}$. The triangle density and hence the clustering coefficient grow considerably faster  than in the previous phase, increasingly so for larger degrees $q$.
    \item {\em Large} $\alpha$: {\em disconnected phase}
    \\[0.5mm] Here the graphs break down completely into large collections of those cliques that had started to appear in the previous phase. The resulting configurations correspond to $q$ regular graphs with the maximum possible number of triangles. In analogy with physics, we call these ground states.
\end{itemize}
We label the transition points between the phases $\alpha_1(N)$ and $\alpha_2(N)$, see  figure \ref{fig:graphs1}. Both $\alpha_1(N)$ and $\alpha_2(N)$ grow logarithmically with $N$. We refer to the transition from connected to partially connected as the \emph{shattering transition} to highlight its topological nature. In figure \ref{fig:graphs2} we show typical graphs sampled via MCMC in the three phases.
\vsp

In order to complement the previous distinctions with quantitative estimates, we will next give an alternative derivation of (\ref{eq:triangledensity}) that  incorporates higher order effects. 
We can always write averages over (\ref{eq:ensemble}) in terms of averages over random regular graphs (RRG), described by the ensemble $p_0(\bA)={\cal N}_q^{-1}\prod_{i\leq N}\delta_{q,\sum_j A_{ij}}$ with uniform probabilities. We will write average over the unbiased random regular graph ensemble $p_0(\bA)$ as $\bra \ldots\ket_{\rm RRG}$.   In particular, 
\begin{eqnarray}
\phi(\alpha)&=& N^{-1}\log\Big( \sum_{\bA\in G}\rme^{\alpha {\rm Tr}(\bA^3)}\prod_{i=1}^N \delta_{q,\sum_j A_{ij}}\Big)
\nonumber
\\
&=& N^{-1}\log\big\bra \rme^{\alpha {\rm Tr}(\bA^3)}\big\ket_{\rm RRG}+N^{-1}\log {\cal N}_q.
\label{eq:generator_approx}
\end{eqnarray}
 We can therefore use some of the rigorous results from random graph theory \cite{bollobas1980probabilistic} established for unbiased ensembles. For instance, a standard result on RRGs \cite{wormald1981asymptotic} concerns the asymptotic  distribution of isolated triangles:
\begin{eqnarray}
\hspace*{-10mm}
    p_N(\triangle) = \big\bra \delta_{\triangle,\triangle(\bA)}\big\ket_{\rm RRG}~\xrightarrow[N\to\infty]{} ~\textrm{Poiss}(\triangle,\lambda),~~
    \lambda= \frac{1}{6}(q\!-\!1)^3,
\end{eqnarray}
in which $\triangle(\bA)$ is the number of isolated triangles in graph $\bA$. 
To understand the shattering transition, we also need to know the statistics of cliques $K_{q+1}$ in RRGs. We do this by splitting the triangle count in two contributions, one from the isolated triangles inside the giant component and the other from the triangles in the cliques $K_{q+1}$. We denote the numbers of each type by $T(\bA)$ and $K(\bA)$ respectively.\footnote{We hereby disregard as insufficiently relevant those triangles that are neither isolated, nor in cliques. This step, which is motivated by what is observed in the MCMC simulations, will be validated a posteriori by the accuracy of the resulting prediction.} Thus
\begin{eqnarray}
    \Tr(\bA^3) = 6 T(\bA) + (q\!+\!1) q (q\!-\!1) K(\bA).
\end{eqnarray}
We now assume that also $K(\bA)$ follows a Poisson distribution, and that for large $N$ the joint distribution of $T$ and $K$ factorizes asymptotically. This assumption  is  to be confirmed  a posteriori in MCMC simulations. Now
\begin{eqnarray}
\hspace*{-10mm}
    p_N(T,K) = \big\bra \delta_{T,T(\bA)}\delta_{K,K(\bA)}\big\ket_{\rm RRG}\approx \textrm{Poiss}(T,\lambda_T) \textrm{Poiss}(K,\lambda_K). 
\end{eqnarray}
Here $\lambda_T=\frac{1}{6}(q\!-\!1)^3$, and 
we can obtain the parameter $\lambda_K$ for the Poisson distribution for the cliques $K_{q+1}$ from \cite{gao2008distribution}. Details are given in \ref{app:subgraphsRRG}. Upon approximating $N!/(N\!-\!q\!-\!1)!\approx N^{q+1}$ we obtain
\begin{eqnarray}
    \lambda_K = \frac{1}{N^{\frac{1}{2}(q-2)(q+1)}}\frac{(q!)^{q+1}}{q^{\frac{q}{2}(q+1)}(q+1)!}.
\end{eqnarray}
We can now proceed
 with the evaluation of (\ref{eq:generator_approx}):
 \begin{eqnarray}
 \hspace*{-10mm}
\phi(\alpha)&\approx & \frac{1}{N}\Big[\log\sum_{T\geq 0}\textrm{Poiss}(T,\lambda_T) \rme^{6\alpha  T}
+\log\sum_{K\geq 0}\textrm{Poiss}(K,\lambda_K)   \rme^{ \alpha  q (q^2-1) }
\nonumber
\\[-3mm]
\hspace*{-10mm}&&\hspace*{80mm} 
+\log {\cal N}_q\Big]
\nonumber
\\[-0.5mm]
&=& \frac{1}{N}\big[\lambda_T\rme^{6\alpha}+\lambda_K\rme^{\alpha q (q^2-1)}-\lambda_T-\lambda_K
+\log {\cal N}_q\big].
\end{eqnarray}
From this directly follows $m(\alpha)=\partial \phi(\alpha)/\partial\alpha$:
\begin{eqnarray}
\label{eq:fullTriangleDensity}
    m(\alpha) &\approx & 
     \frac{1}{N}(q\!-\!1)^3\rme^{6\alpha}+\frac{1}{N^{1+\frac{1}{2}(q-2)(q+1)}}  \frac{(q!)^{q+1}}{q^{\frac{q}{2}(q+1)}(q\!-\!2)!}\rme^{\alpha q (q^2\!-1)}.
     \label{eq:approx_m_alpha}
\end{eqnarray}
The first term coincides with formula (\ref{eq:triangledensity}) from the previous section. The second term in (\ref{eq:fullTriangleDensity}) represents the impact of cliques, and, according to figure \ref{fig:triangleDensity}, accounts for most of the deviations from (\ref{eq:triangledensity}). In spite of our approximation of only accounting for isolated loops and isolated cliques,  the resulting description is seen to  give very good agreement with simulations for the whole range of $\alpha$ values.

As one might expect, the MCMC sampling algorithm slows down as it approaches the ground state. While we will not carry out a detailed dynamical analysis, we will mention the MCMC process slows down considerably precisely in the partially connected phase. To obtain good (i.e. sufficiently independent) samples even close to the ground state, we increased the number of accepted swaps per link in between samples beyond $\alpha_1(N)$ to values in the range  of $10^4-10^5$ accepted swaps per link. We also increased considerably the waiting time before the first sample to $\sim 10^8$ accepted swaps per link, to allow  the system to escape from possible metastable states.

Expression (\ref{eq:fullTriangleDensity}) has a clear interpretation: the expected number of subgraphs of the types $T$ and $K$ are boosted independently when increasing $\alpha$, each with an exponential factor in accordance with the model (\ref{eq:ensemble}). This already provides an explanation for the phases described previously. If we denote by $m_T(\alpha)$ and $m_K(\alpha)$ the first and second term of (\ref{eq:fullTriangleDensity}), then we can describe the phases in terms of the relation between these two terms. 
\begin{itemize}
    \item {\em Small} $\alpha$: {\em connected phase}
    \\[0.5mm]
   Here we have $m_T(\alpha)\gg m_K(\alpha)$. Even though cliques $K_{q+1}$ may be present, their probability is too small to be relevant. 
       \item {\em Intermediate} $\alpha$: {\em partially connected phase}
       \\[0.5mm]
   Here $m_K(\alpha)$ becomes significant. We may define the onset $\alpha_1(N)$ of the partially connected phase to be the point where $m_K(\alpha)=\eta m_T(\alpha)$ for some finite $\eta\in(0,1)$. Here we chose $\eta = 1/10$, which was found appropriate   in the ranges $q\le 6$ and $N\le2000$.  The shattering transition point is then given by
    \begin{eqnarray}
        \alpha_1(N) =& \frac{1}{2}\frac{(q\!-\!2)(q\!+\!1)}{q(q^2\!-\!1)\!-\!6}\log N \nonumber\\
        & + \frac{1}{q(q^2\!-\!1)\!-\!6} \log \p{\eta \frac{(q\!-\!1)^2 q^{\frac{q}{2}(q+1)-1}}{  (q!)^{q}}}.
    \end{eqnarray}
    \item {\em Large} $\alpha$: {\em disconnected phase}
    \\[0.5mm]
   Here the contribution from $m_K(\alpha)$ dominates, and  the whole graph is made of disconnected cliques. The critical value $\alpha_2(N)$ marking the start of this phase is defined by the instance where $m(\alpha_2(N))=q(q\!-\!1)$, i.e. where  the maximum possible density of loops is achieved. We can replace $m(\alpha_2(N))$ by $m_K(\alpha_2(N))$ since the contribution from disconnected triangles is now very small, giving
    \begin{eqnarray}
        \alpha_2 (N) = \frac{1}{2}\frac{(q\!-\!2)(q\!+\!1)\!+\!2}{q(q^2\!-\!1)}\log(N) + \frac{ \log(q)}{2(q\!-\!1)} - \frac{\log(q!)}{q^2\!-\!1}.
    \end{eqnarray}
\end{itemize}
Even though $m_K(\alpha)$ in (\ref{eq:approx_m_alpha}) has a higher power of $1/N$ compared to $m_K(\alpha)$, the prefactor of $\alpha$ for $m_K(\alpha)$ is higher in the exponential, viz. $(q\!+\!1)q(q\!-\!2)>6$ for $q \ge 3$. This means that for any $N$, the clique contribution $m_K(\alpha)$ will always grow faster with $\alpha$ than  $m_T(\alpha)$, which implies it will always surpass it for large enough $\alpha$. We do  not claim that graphs will only be made either of isolated loops in a giant component, but only that knowing the behaviour of the two quantities $m_T(\alpha)$  and $m_K(\alpha)$ appears sufficient to understand the overall behaviour of the loopy ensemble (\ref{eq:ensemble}). 
\vsp

\begin{figure}[t]
    \centering
    \begin{picture}(300,202)
    \put(5,0){\includegraphics[width=0.7\textwidth]{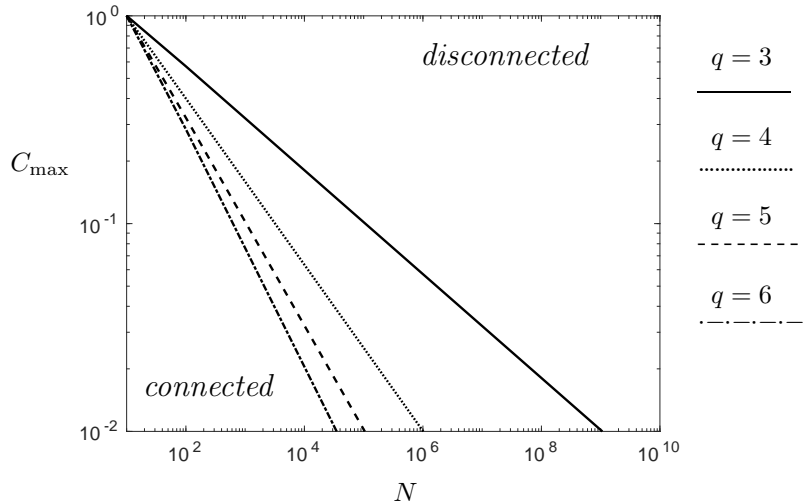}}
    \put(150,160){\large\textrm{\em disconnected}}
    \put(45,35){\large\textrm{\em connected}}
    \put(140,-5){$N$}
    \put(-5,120){$C_{\rm max}$}
    \put(260,160){$q=3$}
    \put(255,150){\line(1,0){35}}
    \put(260,130){$q=4$}
    \multiput(255,120)(2,0){18}{.}
    \put(260,100){$q=5$}
    \multiput(255,90)(5,0){8}{-}
    \put(260,70){$q=6$}
    \multiput(255,60)(10,0){4}{$\cdot-$}
    
    \end{picture}\vspace*{3mm}
    
    \caption{\label{fig:phaseDiagram}The upper bound $C_{\rm max}$ on the  tuneable level of clustering within the giant component, for graphs from the ensemble (\ref{eq:ensemble}), plotted against the graph size $N$. This is shown for different degrees of the regular random graphs. Values for clustering above the lines cannot be achieved in the connected phase, but would require the formation of isolated cliques.} 
\end{figure}

Let us briefly discuss the potential practical utility of (\ref{eq:ensemble}) in light of the previous results. Our ensemble is the maximally unbiased random graph ensemble over regular graphs that satisfies the condition of having a particular clustering coefficient $C$. In order to achieve one's desired value of $C$ it is only necessary to find the appropriate $\alpha(C)$ by solving $C=C(\alpha)= m(\alpha)/q(q\!-\!1)$ using (\ref{eq:fullTriangleDensity}). However, if one's  interest is in using (\ref{eq:ensemble}) as a null model for real networks with link clustering, the presence of cliques is undesirable. If we aim to generate graphs with a single component and a nontrivial number of loops, we need to stay in the connected phase. Moreover,  in this phase we have a very accurate control of $m(\alpha)$ and the spectral density through (\ref{eq:triangle_spectrum}). We conclude that (\ref{eq:ensemble}) can be a useful null model when $C\in (0,C(\alpha_1(N)))$. In that clustering range we can simply take
\begin{eqnarray}
    \alpha(C) = \frac{1}{6}\log\big(N qC/(q\!-\!1)^2\big).
\end{eqnarray}
While the shattering transition occurs at $\alpha=\alpha_1(N)$, the finite size nature of the problem makes it possible that some cliques appear already somewhat earlier. However, it is clear from Figure \ref{fig:graphs1} that an upper bound to the level of clustering achievable without cliques  is given by $C_{\rm max}= C_T(\alpha_2(N))$, the contribution to clustering from disconnected triangles at the transition point to the disconnected phase. 
Values $C \ge C_{\rm max}$ are  impossible to achieve in the ensemble (\ref{eq:ensemble}) without triangles appearing outside the giant component, and additional constraints would have to be introduced into the model to  prevent the formation of isolated cliques. This dependence  on $N$ of the hard upper bound, shown also in Figure \ref{fig:phaseDiagram}, is somewhat unexpected. In particular,
\begin{eqnarray}
    \lim_{N\to\infty} C_{\rm max}= \lim_{N\to\infty}  C_T(\alpha_2(N))=0.
\end{eqnarray}
Hence  for very large sizes, $N\gg 1$, even very small clustering coefficients  are not accessible in the connected regime. This can be understood intuitively as an entropic effect. For fixed value of $C(\alpha)$ and large enough $N$, there are simply many more graphs that achieve $C(\bA)=C(\alpha)$ through cliques than graphs that achieve it through loops embedded in the giant component.

\subsection{Other ensembles}

So far we have focused specifically on the ensemble (\ref{eq:ensemble}) as the simplest nontrivial instance of the more general family (\ref{eq:spectralEnsemble}), suitable for testing limits and for developing further our intuition for the phenomenology of `loopy' random graph ensembles.
However, we have the more general results  (\ref{eq:integral_arbitrary_rhohat},\ref{eq:spectrum_with_J}), applicable to any functional Lagrange parameter  $\hrho[\mu]$, with the key integral expressed as an expansion in Chebyshev polynomials. 
We will now turn to other choices for  $\hrho[\mu]$.\vsp

\begin{figure}[t]
    
    \begin{picture}(372,145)
        \put(20,5){\includegraphics[width=0.5\textwidth,height=0.39\textwidth]{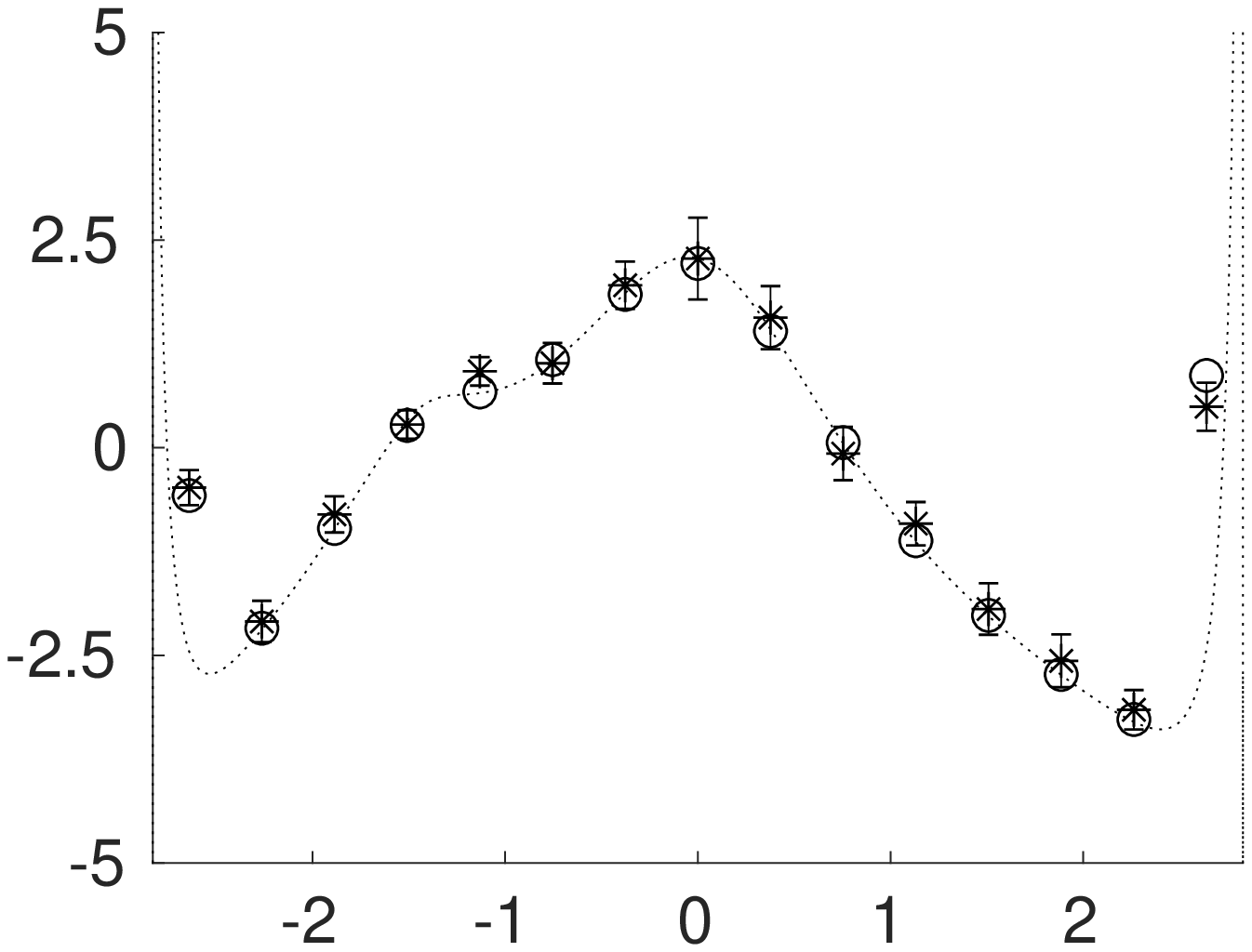}}
        \put(190,5){\includegraphics[width=0.5\textwidth,height=0.39\textwidth]{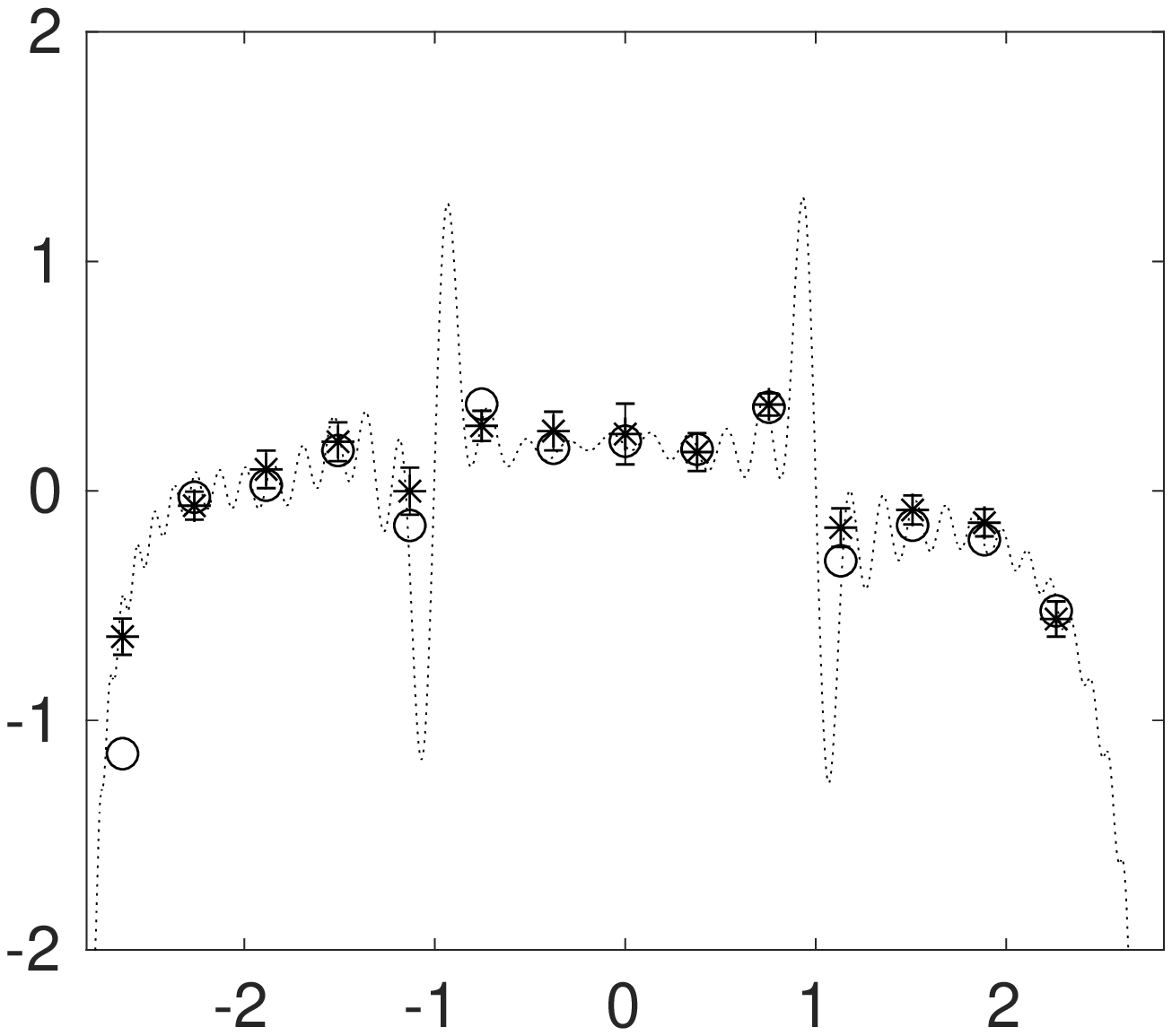}}
        \put(113,-3){$\mu$}
        \put(283,-3){$\mu$}
        \put(0,79){$\delta\varrho(\mu)$}
        \linethickness{.04mm}
        \put(44,21){\line(0,1){118}}
        \linethickness{0.04mm}
        \put(44,139){\line(1,0){144}}
    \end{picture}\vspace*{2mm}
    
    \caption{\label{fig:otherModels} Average spectral densities for more complicated spectrally constrained $q$-regular graph ensembles. 
   As before we show the rescaled finite size deviations from the Kersten-MacKay law  $\delta\varrho(\mu) = N[\varrho(\mu)-\varrho_0(\mu)]$. Left: 
    results for the ensemble (\ref{eq:mixed}), with $q=3$, $N=2000$, and $\alpha = \beta =0.2$. Right: results for the ensemble (\ref{eq:block}), with  $q=3$, $N=1000$, and $\alpha=0.5$. Markers show the average density computed from from histograms of samples obtained via MCMC simulations. The dotted line shows the theoretical predictions, circles show the density prediction for the exact bins as those used for the histograms of the numerical samples.  See the main text for further details. }
\end{figure}

Our first choice is $\hat{\varrho}[\mu]=\alpha\mu^3 +\beta\mu^4$, which generalizes the ensemble (\ref{eq:ensemble}) in that we now control the number of closed paths of both length 3 and length 4:
\begin{eqnarray}
    p(\bA) &=& \frac{\rme^{\alpha \Tr (\bA^3)+\beta\Tr(\bA^4) }}{Z(\alpha,\beta)}\prod_{i=1}^N\delta_{q,\sum_j A_{ij}},
      \label{eq:modelMixed}
    \\
    Z(\alpha,\beta)&=&\sum_{\bA\in G}\rme^{\alpha \Tr (\bA^3)+\beta\Tr(\bA^4) }\prod_{i=1}^N\delta_{q,\sum_j A_{ij}}.
    \end{eqnarray}
    The calculations for this ensemble (\ref{eq:modelMixed}) are very similar to those carried out for (\ref{eq:ensemble}), which allows us to be brief. 
  We have already computed  the relevant integrals  in (\ref{eq:Jintegral_triangles},\ref{eq:Jintegral_squares}), and we can therefore immediately proceed to the spectral density:
  \begin{eqnarray}
  \hspace*{-15mm}   \varrho(\mu) &=& \varrho_0(\mu) + \frac{1}{N}\varrho_1(\mu)+ \sum_{\ell=3}^\infty \frac{(q\!-\!1)^\ell}{2N}\Big(\rme^{6\alpha\delta_{3,\ell} + 8\beta \delta_{4,\ell}}\!-\!1\Big)
     [g_\ell(\mu) \!-\! h(\mu)]  +o(\frac{1}{N})
    \nonumber
    \\
   \hspace*{-15mm}     &=& \varrho_0(\mu)+ \frac{1}{N}\varrho_1(\mu) + \frac{(q\!-\!1)^3}{2N}(\rme^{6\alpha}\!-\!1)[g_3(\mu) \!-\! h(\mu)] 
   \nonumber
   \\
   \hspace*{-15mm}&&\hspace*{20mm}
   + \frac{(q\!-\!1)^4}{2N}(\rme^{8\beta}\!-\!1)[g_4(\mu)\! -\! h(\mu)]   +o(N^{-1}),
       \label{eq:mixed}
\end{eqnarray}
in which the functions $g_\ell(\mu)$ and $h(\mu)$ are given in (\ref{eq:hmu_general},\ref{eq:glmu_general}). 
A comparisons of the predicted spectrum (\ref{eq:mixed}) with measurents in MCMC simulations, for $N=2000$ and $\alpha=\beta=0.2$, is shown in the left panel of Figure \ref{fig:otherModels}.
The MCMC algorithm used was similar to the one described before, but now they also require monitoring the evolution of  $\Tr(\bA^4)$ (in addition to $\Tr(\bA^3)$), as both appear in  the move acceptance probabilities. In each run 100 samples were generated from each initial seed, after a burn-in (waiting time) of $ \sim 10^3$ swaps per link. Error bars give the standard deviation corresponding to fluctuations between 10 different initial seeds, so that a total of 1000 graphs were averaged. As in the previous case, we recover the results from \cite{lopez2018exactly} when setting $q=2$. As one would expect, (\ref{eq:mixed}) is only valid in the vicinity of $(\alpha,\beta)=(0,0)$, to avoid the emergence of extensively many small fully connected $q$-regular bipartite graphlets, which maximize the number of $4-$loops around a node.
\vsp

Our second alternative choice for $\hat{\varrho}[\mu]$ is the following block function, 
which  introduces a bias in the graph probabilities depending on the number of eigenvalues inside the interval $[-1,1]$:
\begin{eqnarray}
    \label{eq:modelBlock}
    \hrho(\mu) &= \alpha~ \theta(1-\abs{\mu}). 
\end{eqnarray}
Now we have $N\int\!\rmd\mu~\hat{\varrho}(\mu)\varrho(\mu|\bA) =\alpha \mathcal{I}(\bA |[-1,1])$, where 
 $\mathcal{I}(\bA|[-1,1])$ denotes the number of eigenvalues of $\bA$ inside the interval $[-1,1]$. 
In contrast to powers of $\mu$, understanding intuitively the topological effects of the choice (\ref{eq:modelBlock}) is not straightforward, notwithstanding the clear nontrivial effect on the observed spectrum. 
In this case we have ${\cal J}_\ell[\hat{\varrho}]=\alpha\int\! \rmd\mu~\theta[1\!-\!|\mu|][g_\ell(\mu) \!-\! h(\mu)]\not = 0$ for all $\ell$, so we introduce a (sufficiently large) cutoff $L$ in the summation of (\ref{eq:spectrum_with_J}). Since $|{\cal J}_\ell[\hat{\varrho}]|\ll 1$ we set this integral to zero for $\ell > L$, as was done previously in \cite{lucibello2014finite}, leaving the truncation
\begin{eqnarray}
\label{eq:block}
    \varrho(\mu)  &=& \rho_0(\mu)+ \frac{1}{N}\rho_1(\mu)
    \nonumber
    \\&&
 + \sum_{\ell=3}^L \frac{(q\!-\!1)^\ell}{2N}\Big[\rme^{\alpha \ell \int_{-1}^1\!\rmd\mu~[g_\ell(\mu)-h(\mu)]}\! -\!1\Big] [g_\ell(\mu) - h(\mu)],
 \end{eqnarray}
 in which the integrals can be worked out in more explicit form, as we did for the previous cases.
In figure \ref{fig:otherModels} (right panel) we compare the prediction (\ref{eq:block})  with results from numerical  MCMC samples,  and observe a good agreement. We point out that generating graph samples from the spectrally constrained ensemble (\ref{eq:spectralEnsemble}, \ref{eq:modelBlock}) numerically is considerably more computationally expensive than the for the previous models. Here, instead of the number of triangles or squares, the number of eigenvalues inside the interval $[-1,1]$ has to be monitored. This requires that the full set of eigenvalues of the graph $\bA$ has to be calculated after \emph{each} edge swap, which necessitated  parallel execution in multi-core computers, to reduce CPU  time to a few weeks.  We seen in Figure \ref{fig:otherModels} that the deviations from $\varrho_0(\mu)$ are quite small, nevertheless they are nontrivial and are predicted accurately. To measure spectra at this level of detail, we averaged over $10^4$ graphs, separated during the MCMC process by $\sim 1$ swaps per link. Error bars are obtained by splitting this data set in groups of $10$. 

The above results are similar in form to the ones derived for weighted graphs in \cite{castillo2018theory}, the main difference is that in  \cite{castillo2018theory} a second set of replicas with the traditional limit $n\to 0$ is introduced to get the spectrum. It is interesting to see  that with the functional formalism (\ref{eq:spectralEnsemble}) both the observable $\int\!\rmd\mu~\hrho(\mu)\varrho(\mu)$ and the spectrum $\varrho(\mu)$ itself are calculated at the same time.

\section{Discussion}

In this paper we have extended and applied an analytic approach for describing constrained maximum entropy ensembles of  finitely connected random loopy graphs of large but finite size. We focused on regular random graphs with  soft constrained adjacency matrix eigenvalue spectra. We were able to develop a general theory describing the $\Order{1/N}$ contributions to the expected eigenvalue spectrum, through the use of an infinite number of replica indices taking values in the imaginary axis \cite{CoolenLoopy}, and building on techniques from earlier studies such as \cite{kuhn2008spectra,metz2014finite,lucibello2014finite,castillo2018theory}.

The simplest nontrivial spectrally constrained ensembles are those in which the spectral constraint reduces to a soft constraint on the number of triangles.  We quantified the behaviour of such systems, which following \cite{lopez2018exactly} we have come to regard as the `harmonic oscillators' of loopy graph ensembles, and showed how they allow for fine tuning of their average clustering coefficients. A limitation on their use as null models for regular graphs with nontrivial clustering is that there is a maximum achievable clustering coefficient, whose value depends on the size of the graph,  beyond which the ensemble undergoes a transition into a new phase, where high clustering levels are achieved by the graph fracturing into extensively many disconnected cliques. We presented a precise analytic estimate for an upper bound on the maximum clustering coefficient that is achievable without fracturing of the graph. We also showed how the general theory applies to other spectrally constrained ensembles, such as those where both the number of triangles and the number of squares are controlled, and to ensembles where the spectral constraint reduces to a count of the number of adjacency matrix eigenvalues in a given interval.  We  carried out numerical simulations via MCMC processes based on edge swaps with nontrivial acceptance probabilities, which are themselves generally nontrivial in view of the need to recompute eigenvalue spectra after each accepted move. In comparing triangle counts and spectra, we found excellent agreement between the theoretical predictions and the MCMC measurements in all cases, provided we remain in the parameter regime where higher orders in $N$ of the generating function are not yet important.

The most natural generalization of the presently studied family of models would be to extend the imaginary replica approach to sparse graphs with an arbitrary degree distribution $p(k)$. Preliminary numerical simulations show that these non-regular graph ensembles  behave in a very similar way to what has been observed for regular graphs. In addition it would be interesting to explore further the possibility of controlling short loops in finitely connected graphs without this being realized microscopically by such graphs fracturing into extensively many disconnected graphlets, even at high loop densities. This would seem to require more complicated choices of the functional Lagrange parameter $\hrho(\mu)$ than the ones studied so far, possibly including choices that scale differently with $N$. Both these directions for further research are now being explored by the authors. 
\\[3mm]
{\bf Acknowledgement}
\\[1mm]
FAL gratefully acknowledges financial support through a scholarship from Conacyt (Mexico). The authors are grateful to  Alexander Mozeika and Diego Vidal Cruzprieto for very valuable discussions.

\section*{References}

\providecommand{\newblock}{}

\appendix
\section{The functional integral}
\label{appendix:changeVariable}

\subsection{Transformation to Fourier components}

In this Appendix we simplify expression (\ref{eq:generator_in_nasty_integral},\ref{eq:nasty_functional_integral}) for our generating function $\phi[\hat{\varrho}]$. 
We first introduce a number of definitions to compactify our formulae:
\begin{eqnarray}
&& \hspace*{-5mm}   P(\bvarphi,\omega)= \sum_{\ell\in\mathbb{Z}} W_\ell(\bvarphi) \frac{\rme^{\rmi\ell\omega}}{2\pi},~~~~~~~
     W_\ell(\bvarphi) = \int_{0}^{2\pi}\!\rmd\omega~ P(\bvarphi,\omega)\rme^{-\rmi\ell\omega},
     \\[-1mm]
     && \hspace*{-5mm} 
    \hP(\bvarphi,\omega) = \sum_{\ell\in\mathbb{Z}}\hW_\ell(\bvarphi)\rme^{-\rmi\ell\omega},~~~~~~
    \hW_\ell(\bvarphi) = \int_0^{2\pi}\!\frac{\rmd\omega}{2\pi} ~\hP(\bvarphi,\omega)\rme^{\rmi\ell\omega}.
\end{eqnarray}
We then write  $S[P,\hP] $ strictly in terms of the Fourier transforms $\{W_\ell(\bvarphi),\hat{W}_\ell(\bvarphi)\}$, noting that $\omega\in[-\pi,\pi]$. The result is
\begin{eqnarray}
S[P,\hat{P}]=S[W_1,\hat{W}_1,W_2,\hat{W}_2]+S_0[\{W_\ell,\hat{W}_\ell\}],
\end{eqnarray}
where, using the notation $\berr=\{r_\ell\geq 0,~\ell\in \mathbb{Z}\}$,
    \begin{eqnarray}
    \hspace*{-15mm}
      S_0[\{W_\ell,\hW_\ell\}] &=&    
      \rmi \sum_{\ell\notin\{1,2\}}\int\!\rmd\bvarphi  ~W_\ell(\bvarphi)  \hW_\ell(\bvarphi)     
      \nonumber
      \\[-1mm]
        \hspace*{-15mm}
     &&  + \log \!\int \! \rmd\bvarphi~ \nu(\bvarphi) \sum_{\berr}
   \delta_{q,\sum_{\ell\in\mathbb{Z}}\ell r_\ell}
   \prod_{\ell\in\mathbb{Z}}
  \frac{[- \rmi   \hW_\ell(\bvarphi)]^{r_\ell}}{r_\ell!},
      \\
        \hspace*{-15mm}
    S[W_1,\hW_1, W_2, \hW_2] &=&
     -\frac{q}{2}(1\!+\! \frac{q\!-\!2}{2N})
    + \rmi \int\!\rmd\bvarphi  \Big[W_1(\bvarphi)  \hW_1(\bvarphi)   \!+ \!W_2(\bvarphi)  \hW_2(\bvarphi)     \Big]
     \nonumber
     \\
     \hspace*{-15mm}&&
     \hspace*{-18mm}
     - \frac{q}{2N}\!  \int\!\rmd\bvarphi~W_2(\bvarphi) 
     \rme^{-\rmi \bvarphi \cdot \bvarphi} 
  + \frac{q}{2}(1\!+\!\frac{q}{N})
\! \int\!\rmd\bvarphi\rmd\bvarphi ~W_1(\bvarphi)W_1(\bvarphi^\prime)\rme^{-\rmi \bvarphi \cdot \bvarphi^\prime }
 \nonumber
 \\[-2mm]
 \hspace*{-15mm} &&
-\frac{q^2}{4N}\! \int\!\rmd\bvarphi\rmd\bvarphi~
W_2(\bvarphi)
W_{2}(\bvarphi^\prime) 
\rme^{-2\rmi \bvarphi \cdot \bvarphi^\prime } . 
  \end{eqnarray}
  The integration in (\ref{eq:generator_in_nasty_integral})  can be replaced by 
 integration over the functional Fourier components, since (apart from an irrelevant multiplicative constant) the Jacobian of this coordinate transformation is unitary. We define  $\calD W= \prod_{\bvarphi}[ \rmd W(\bvarphi) \sqrt{N\Delta_{\varphi}/2\pi}]$, and the functional delta distribution $\delta[W]=\prod_{\bvarphi}[\delta(W(\bvarphi))\sqrt{2\pi/N\Delta_\varphi}]$, where $\delta(x)$ is the ordinary delta distribution, so that for any smooth $F[W]$ we will have:
 \begin{eqnarray}
\hspace*{-10mm} \int\!\calD W~F[W]\delta[W]=F[0],~~~~~~
\delta[W]= \int\!\calD \hat{W}~
\rme^{\rmi N\!\int\!\rmd\bvarphi  ~W(\bvarphi)  \hW(\bvarphi) ]},
\end{eqnarray}
and the generating function can be written as, modulo an irrelevant additive constant:
\begin{eqnarray}
\hspace*{-10mm}
\phi[\hat{\varrho}]&=& \lim\frac{1}{N}\log \int\!\Big[\prod_{\ell\in\Z}\calD W_\ell \calD \hW_{\ell}\Big]
\rme^{NS[W_1,\hW_1,W_2,\hW_2]+NS_0[\{W_\ell,\hW_\ell\}]}.
\end{eqnarray}
With (A.6), integration over the Fourier components with $\ell=1,2$ has become trivial:
\begin{eqnarray}
&&\hspace*{-25mm}
\int\!\!\prod_{\ell\notin\{1,2\}}\![\calD W_\ell \calD \hW_\ell]~\rme^{NS_0[\{W_\ell,\hW_\ell\}] }
= \rme^{\rmi N\sum_{\ell=1}^2\int\!\rmd\bvarphi  ~W_\ell(\bvarphi)  \hW_\ell(\bvarphi) }\nonumber
\\[-2.5mm]
&&\hspace*{-12mm}\times\!\int\!\!\!\prod_{\ell\notin\{1,2\}}\!\!\big[\calD \hW_\ell~ \delta[\hW_\ell]\big]
\exp\Big( N\log \!\int \! \rmd\bvarphi~ \nu(\bvarphi) \sum_{\berr}
   \delta_{q,\sum_{\ell\in\mathbb{Z}}\ell r_\ell}
   \prod_{\ell\in\mathbb{Z}}
  \frac{[- \rmi   \hW_\ell(\bvarphi)]^{r_\ell}}{r_\ell!}\Big)
  \nonumber
   \\
  && \hspace*{-10mm} =
   \exp\Big(\rmi N\sum_{\ell=1}^2\int\!\rmd\bvarphi  ~W_\ell(\bvarphi)  \hW_\ell(\bvarphi) 
   \nonumber
   \\[-2mm]
   &&\hspace*{5mm}
 +N\log \!\int \! \rmd\bvarphi~ \nu(\bvarphi) \sum_{0\leq r\leq q/2}
  \frac{[- \rmi   \hW_1(\bvarphi)]^{q-2r}}{(q-2r)!}\frac{[- \rmi   \hW_2(\bvarphi)]^{r}}{r!}
   \Big)\Big].
\end{eqnarray}
Hence (\ref{eq:generator_in_nasty_integral})  can be written as follows:
\begin{eqnarray}
\hspace*{-20mm}
\phi[\hat{\varrho}]&=& -\frac{q}{2}(1\!+\! \frac{q\!-\!2}{2N})
+ \lim \frac{1}{N}\log \int\!
\calD \hW_1 \calD \hW_2
\exp\Big\{N\Big(
S_1[\hat{W}_1] + S_2[\hat{W}_2] 
\nonumber
\\
\hspace*{-15mm}
&&\hspace*{15mm} +~
N\log\! \!\int \! \rmd\bvarphi~ \nu(\bvarphi)\! \!\sum_{0\leq r\leq q/2}\!\!
  \frac{[- \rmi   \hW_1(\bvarphi)]^{q-2r}}{(q-2r)!}\frac{[- \rmi   \hW_2(\bvarphi)]^{r}}{r!}
  \Big)\Big\},
  \label{eq:generator_hats_only}
  \end{eqnarray}
  with the following two functionals $S_{1,2}[\hat{W}]$:
  \begin{eqnarray}
  \hspace*{-10mm}
  S_1[\hat{W}]&=&\frac{1}{N}\log 
 \int\! \calD W \exp\Big(
     \rmi N\!\int\!\rmd\bvarphi  ~W(\bvarphi)  \hW(\bvarphi) 
     \nonumber
     \\
     \hspace*{-10mm}&&
  \hspace*{20mm}+ \frac{q}{2}(N\!+\!q)
\! \int\!\rmd\bvarphi\rmd\bvarphi ~W(\bvarphi)U_1(\bvarphi,\bvarphi^\prime)W(\bvarphi^\prime)
  \Big)
  \label{eq:S1}
  \\
   \hspace*{-10mm}
  S_2[\hat{W}]&=&
  \frac{1}{N}\log 
 \int\!  \calD W \exp\Big(
     \rmi N\!\int\!\rmd\bvarphi  ~W(\bvarphi)  \hW(\bvarphi)   
     - \frac{q}{2}\!  \int\!\rmd\bvarphi~W(\bvarphi) 
    V(\bvarphi) 
     \nonumber
     \\
     \hspace*{-10mm}
     &&
      \hspace*{20mm}
-\frac{1}{4}q^2 \int\!\rmd\bvarphi\rmd\bvarphi~
W(\bvarphi)U_2(\bvarphi,\bvarphi^\prime)
W(\bvarphi^\prime) 
  \Big),
    \label{eq:S2}
\end{eqnarray}
where we used the short-hands $U_n(\bvarphi,\bvarphi^\prime)=\exp(-\rmi n\bvarphi\cdot\bvarphi^\prime)$ and $V(\bvarphi)=U_1(\bvarphi,\bvarphi)$.

\subsection{Gaussian functional integrals}

Both $ S_1[\hat{W}]$ and $ S_2[\hat{W}]$ involve complex Gaussian functional integrals, of the following form, with $d={\rm dim}(\bvarphi)$ and $\Delta_\varphi\to 0$ in the functional integration limit:
\begin{eqnarray}
\hspace*{-10mm}
J[U,Q]&=&
 \int\! \calD W  ~\rme^{-\frac{1}{2}\int\!\rmd\bvarphi\rmd\bvarphi^\prime ~W(\bvarphi)U(\bvarphi,\bvarphi^\prime)W(\bvarphi^\prime)
 + \int\!\rmd\bvarphi ~W(\bvarphi)Q(\bvarphi)}
 \nonumber
 \\
 \hspace*{-10mm}
 &=&
  \int\! \prod_{\bvarphi }\frac{\rmd W(\bvarphi)}{\sqrt{2\pi/N\Delta_\varphi}} ~\rme^{-\frac{1}{2}\Delta_\varphi^2\sum_{\bvarphi \bvarphi^\prime} W(\bvarphi)U(\bvarphi,\bvarphi^\prime) W(\bvarphi^\prime)
 +\Delta_\varphi\sum_{\bvarphi} W(\bvarphi)Q(\bvarphi)}
 \nonumber
 \\
  \hspace*{-10mm}
 &=& 
\Big(\frac{N}{\Delta_\varphi}\Big)^{d/2} \!\!\frac{1}{\sqrt{{\rm Det} \bU}}~\rme^{\frac{1}{2}\Delta_\varphi^2\sum_{\bvarphi \bvarphi^\prime} (\Delta_\varphi^2\bU)^{-1}(\bvarphi,\bvarphi^\prime) Q(\bvarphi)Q(\bvarphi^\prime)}.
\label{eq:general_Gaussian}
\end{eqnarray}
$\bU$ is the  matrix of discretized values $U(\bvarphi,\bvarphi^\prime)$. The entries of the inverse functional kernel $U^{-1}(\bvarphi,\bvarphi^\prime)$, defined by  the condition
$\delta(\bvarphi\!-\!\bvarphi^\prime)= \int\!\rmd\bvarphi^\pprime ~U^{-1}(\bvarphi,\bvarphi^\pprime)U(\bvarphi^\pprime,\bvarphi^\prime)$, are $U^{-1}(\bvarphi,\bvarphi^\prime)=(\Delta^2_\varphi\bU)^{-1}(\bvarphi,\bvarphi^\prime)$. Hence the following identities hold:
\begin{eqnarray}
J[U,Q]&=&
\Big(\frac{N}{\Delta_\varphi}\Big)^{d/2}\!\! \frac{1}{\sqrt{{\rm Det} \bU}}~\rme^{\frac{1}{2}\!\int\!\rmd \bvarphi \rmd \bvarphi^\prime Q(\bvarphi) U^{-1}(\bvarphi,\bvarphi^\prime) Q(\bvarphi^\prime)},
\label{eq:general_Gaussian_result}
\\
J[U^{-1}\!,Q]&=&
(N\Delta_\varphi)^{d/2}~\sqrt{{\rm Det} \bU}~\rme^{\frac{1}{2}\!\int\!\rmd \bvarphi \rmd \bvarphi^\prime Q(\bvarphi) U(\bvarphi,\bvarphi^\prime) Q(\bvarphi^\prime)}.
\end{eqnarray}
We can now work out (\ref{eq:S1}) and (\ref{eq:S2}), and find
 \begin{eqnarray}
  \hspace*{-20mm}
 \rme^{N S_1[\hat{W}]}&=& J\big[\!-\!q(N\!+\!q)U_1,\rmi N\hW\big]
 \nonumber
 \\
   \hspace*{-20mm}
 &=&\Big(\frac{N}{-q(N+q)\Delta_\varphi}\Big)^{\!d/2}\!\!\frac{1}{\sqrt{{\rm Det} \bU_1}}~
\rme^{\frac{N^2}{2q(N+q)}\!\int\!\rmd \bvarphi \rmd \bvarphi^\prime ~\hW(\bvarphi)  U_1^{-1}(\bvarphi,\bvarphi^\prime)   \hW(\bvarphi^\prime) },
\\
   \hspace*{-20mm}
  \rme^{N  S_2[\hat{W}]}&=& J\big[\frac{1}{2}q^2U_2,\rmi N\hat{W}\!-\!\frac{1}{2}qV\big]
  \nonumber
  \\
  \hspace*{-20mm}&=&
  \Big(\frac{2N}{q^2\Delta_\varphi}\Big)^{\!d/2}
  \!\!\!\frac{1}{\sqrt{{\rm Det} \bU_2}}~
  \rme^{-\frac{1}{q^{2}}\!\int\!\rmd \bvarphi \rmd \bvarphi^\prime [N \hW(\bvarphi)   
     + \frac{q}{2}\rmi V(\bvarphi)] U_2^{-1}(\bvarphi,\bvarphi^\prime) [N \hW(\bvarphi^\prime)   
     + \frac{q}{2}\rmi V(\bvarphi^\prime)]  }.
     \nonumber
     \\[-2mm]
     \hspace*{-20mm}&&
  \end{eqnarray}
  For   (\ref{eq:generator_hats_only}) this implies:
  \begin{eqnarray}
\hspace*{-10mm}
\phi[\hat{\varrho}]&=& -\frac{q}{2}(1\!+\! \frac{q\!-\!2}{2N})
+ \lim \frac{1}{N}\log
\Big(\frac{2N}{q^2\Delta_\varphi}\Big)^{\!d/2}\!\!
\frac{1}{\sqrt{{\rm Det} \bU_2}}\int\!
\calD \hW_1~ \rme^{NS_1[\hW_1]}
\nonumber
\\
\hspace{-10mm}
&&
\times\!\int\!\calD \hW_2~\rme^{
-\frac{1}{q^2}\!\int\!\rmd \bvarphi \rmd \bvarphi^\prime [N\hW_2(\bvarphi)   
     + \frac{q}{2}\rmi V(\bvarphi)] U_2^{-1}(\bvarphi,\bvarphi^\prime) 
     [N\hW_2(\bvarphi^\prime)   
     + \frac{q}{2}\rmi V(\bvarphi^\prime)] }
      \nonumber
     \\
\hspace*{-10mm}
&&\times\exp\Big\{
N\log\! \!\int \! \rmd\bvarphi~ \nu(\bvarphi)\! \!\sum_{0\leq r\leq q/2}\!\!
  \frac{[- \rmi   \hW_1(\bvarphi)]^{q-2r}}{(q-2r)!}\frac{[- \rmi   \hW_2(\bvarphi)]^{r}}{r!}
  \Big\}.
  \end{eqnarray}
We transform $\hat{W}_2(\bvarphi)\to \hat{W}_2(\bvarphi)/N-q\rmi V(\bvarphi) /2N$ and 
and expand for large $N$, following \cite{metz2014finite}.
\footnote{Note that we could also have chosen the transformation $\hat{W}_2(\bvarphi)\to \hat{W}_2(\bvarphi)/\sqrt{N}-q\rmi \rme^{-\rmi \bvarphi \cdot \bvarphi} /2N$, but this would in subsequent stages of the calculation have prompted a further rescaling of $\hat{W}_2$ by $\sqrt{N}$.} This assures that $\phi[\hat{\varrho}]$ remains well-defined and nontrivial, and that the leading orders in $N$ of the generating function can be written as follows (assuming that $q>1)$:
  \begin{eqnarray}
\hspace*{-20mm}
\phi[\hat{\varrho}]&=& -\frac{q}{2}(1\!+\! \frac{q\!-\!2}{2N})
+ \lim \frac{1}{N}\log 
\Big(\frac{2}{Nq^2\Delta_\varphi}\Big)^{\!d/2}
\!\!\frac{1}{\sqrt{{\rm Det} \bU_2}}\int\!
\calD \hW_1 ~\rme^{NS_1[\hW_1]}
\nonumber
\\
\hspace*{-20mm}&&
\times\rme^{
N\log\int \! \rmd\bvarphi~ \nu(\bvarphi)
  [-\rmi \hW_1(\bvarphi)]^q/q!}
 \!\int\!\calD \hW_2~\rme^{
-\frac{1}{2}\!\int\!\rmd \bvarphi \rmd \bvarphi^\prime~\hW_2(\bvarphi) (\frac{1}{2}q^2 U_2)^{-1}(\bvarphi,\bvarphi^\prime)   \hW_2(\bvarphi^\prime)}
\nonumber
\\
\hspace*{-20mm}
&&\hspace*{-5mm}
\times\exp\Big\{\!\!-\!\rmi q(q\!-\!1)\frac{\int\!\rmd\bvarphi~\nu(\bvarphi)[-\rmi\hW_1(\bvarphi)]^{q-2}[\hW_2(\bvarphi)\!-\!\frac{1}{2}\rmi qV(\bvarphi)]}{\int\!\rmd\bvarphi^\prime~\nu(\bvarphi^\prime)[-\rmi\hW_1(\bvarphi^\prime)]^q}+{\mathcal O}(\frac{1}{N})\Big\}.
  \end{eqnarray}
  We can now integrate over $\hW_2$, and 
 with the short-hand
 \begin{eqnarray}
 R[\hat{W}(\bvarphi)]&=&q(q\!-\!1)\frac{ \nu(\bvarphi)
    [-\rmi \hW(\bvarphi)]^{q-2}}{\int \! \rmd\bvarphi^\prime~ \nu(\bvarphi^\prime)
 [-\rmi  \hW(\bvarphi^\prime)]^{q}}
\end{eqnarray}
the result takes the form 
  \begin{eqnarray}
\hspace*{-20mm}
\phi[\hat{\varrho}]&=&
 -\frac{q}{2}(1\!+\! \frac{q\!-\!2}{2N})   + \lim \frac{1}{N}\log  \int\!
\calD \hW_1 ~\rme^{-\frac{1}{4}q^2\!\int\!\rmd \bvarphi \rmd \bvarphi^\prime R[\hat{W}_1(\bvarphi)]  U_2(\bvarphi,\bvarphi^\prime) R[\hat{W}_1(\bvarphi^\prime)] }
\nonumber
\\
\hspace*{-25mm}&& \times ~\rme^{NS_1[\hW_1]+N\log
\int \! \rmd\bvarphi~ \nu(\bvarphi)
  [-\rmi \hW_1(\bvarphi)]^q/q!
- \frac{1}{2}\!\int \! \rmd\bvarphi~R[\hW_1(\bvarphi)]
   V(\bvarphi)+{\mathcal O}(N^{-1})}
   \nonumber
   \\
   \hspace*{-20mm}
   &=&
 -\frac{q}{2}(1\!+\! \frac{q\!-\!2}{2N})   + \lim \frac{1}{N}\log  \int\!
\calD \hW_1 ~\rme^{-\frac{1}{4}q^2\!\int\!\rmd \bvarphi \rmd \bvarphi^\prime R[\hat{W}_1(\bvarphi)]  U_2(\bvarphi,\bvarphi^\prime) R[\hat{W}_1(\bvarphi^\prime)] }
\nonumber
\\
\hspace*{-25mm}&& \times \Big(\frac{N}{-q(N\!+\!q)\Delta_\varphi}\Big)^{\!d/2}\!\frac{1}{\sqrt{{\rm Det}\bU_1}}~\rme^{ \frac{N^2}{2q(N\!+\!q)}\!\int\!\rmd \bvarphi \rmd \bvarphi^\prime~\hW_1(\bvarphi)  U_1^{-1}(\bvarphi,\bvarphi^\prime)   \hW_1(\bvarphi^\prime) }
\nonumber
\\
\hspace*{-20mm}&&
\times\rme^{
N\log
\int \! \rmd\bvarphi~ \nu(\bvarphi)
  [-\rmi \hW_1(\bvarphi)]^q/q!
- \frac{1}{2}q\!\int \! \rmd\bvarphi~R[\hW_1(\bvarphi)]
   V(\bvarphi)+{\mathcal O}(N^{-1})}.
  \end{eqnarray}
 Finally we transform $\hat{W}_1(\bvarphi)=\rmi\int\!\rmd\bvarphi^\prime ~ U_1(\bvarphi,\bvarphi^\prime)W(\bvarphi^\prime)$, which gives apart from irrelevant additive constants:
     \begin{eqnarray}
\hspace*{-5mm}
\phi[\hat{\varrho}]&=&
 \lim \frac{1}{N}\log \Big\{\sqrt{{\rm Det}(q\bU_1)}  \int\!
\calD W ~\rme^{N\calS_0[W]+\calS_1[W]
 +{\cal O}(N^{-1})}\Big\},
 \label{eq:phi_before_saddle}
 \end{eqnarray}
 with  
 \begin{eqnarray}
 \hspace*{-15mm}
 \calS_0[W]&=& 
   -\frac{q}{2}\!\int\!\rmd \bvarphi \rmd \bvarphi^\prime~W(\bvarphi) U_1(\bvarphi,\bvarphi^\prime)W(\bvarphi^\prime) 
   \nonumber
\\
\hspace*{-15mm}&&\hspace*{15mm}
+ \log
\int \! \rmd\bvarphi~ \nu(\bvarphi)
 \Big[\int\!\rmd\bvarphi^\prime~ U_1(\bvarphi,\bvarphi^\prime)W(\bvarphi^\prime)\Big]^q,
\\
 \hspace*{-15mm}
\calS_1[W]&=&  
-\frac{1}{4}(q\!-\!1)^2\!\int\!\rmd \bvarphi \rmd \bvarphi^\prime ~r[W(\bvarphi)] U_2(\bvarphi,\bvarphi^\prime)
 r[W(\bvarphi^\prime)] 
\\
\hspace*{-15mm}
&&
 +\frac{1}{2}q^2\!\int\!\rmd \bvarphi \rmd \bvarphi^\prime~W(\bvarphi) U_1(\bvarphi,\bvarphi^\prime)W(\bvarphi^\prime) 
-
\frac{1}{2}(q\!-\!1)\!\int \! \rmd\bvarphi~r[W(\bvarphi)]
  V(\bvarphi),
  \nonumber
\end{eqnarray}
and
\begin{eqnarray}
\hspace*{-5mm}
r[W(\bvarphi)]=  \frac{ \nu(\bvarphi)
  [\int\rmd\bvarphi^\prime ~ U_1(\bvarphi,\bvarphi^\prime) W(\bvarphi^\prime)]^{q-2}}{\int \! \rmd\bvarphi^\prime~ \nu(\bvarphi^\prime)
  [\int\rmd\bvarphi^\pprime ~U_1(\bvarphi^\prime,\bvarphi^\pprime) W(\bvarphi^\pprime)]^q}.
\end{eqnarray}

\subsection{Leading two orders via saddle point integration}

Expression (\ref{eq:phi_before_saddle}) allows us in the usual manner to calculate the leading two orders
in $N$ of the generating function, by substituting $W = W_0 +N^{-\frac{1}{2}}W_1$, where $W_0$ is the
saddle point of $\calS_0[W]$ and where $W_1 = {\mathcal O}(1)$. We obtain, again modulo a constant:
\begin{eqnarray}
\hspace*{-15mm}
\phi[\hat{\varrho}]&=&\lim\Big\{\calS_0[W_0]+\frac{1}{N}\calS_1[W_0]+\frac{1}{N}\log\Big[\frac{\sqrt{{\rm Det}(q\bU_1)}}{\sqrt{{\rm Det}(-\bGamma)}}\Big]\Big\}+{\mathcal O}(N^{-\frac{3}{2}}),
\end{eqnarray}
in which we have the functional curvature at the saddle point:
\begin{eqnarray}
\Gamma(\bvarphi,\bvarphi^\prime)=\frac{\delta^2\calS[W]}{\delta W(\bvarphi)\delta W(\bvarphi^\prime)}\Big|_{W_0}.
\label{eq:curvature}
\end{eqnarray}
What remains is to compute $W_0(\bvarphi)$ and $\Gamma(\bvarphi,\bvarphi^\prime)$. 
Setting $\delta\calS_0/\delta W=0$, and using
the symmetry of $U_1$, gives the saddle point equation from which to solve $W_0$:
\begin{eqnarray}
W_0(\bvarphi)&=&\frac{1}{Z_q}\nu(\bvarphi)\Big[\int\!\rmd\bvarphi^\prime~U_1(\bvarphi,\bvarphi^\prime)W_0(\bvarphi^\prime)\Big]^{q-1},
\label{eq:spe1}
\\
Z_q&=&\int\!\rmd\bvarphi~ \nu(\bvarphi)\Big[\int\!\rmd\bvarphi^\prime~U_1(\bvarphi,\bvarphi^\prime)W_0(\bvarphi^\prime)\Big]^{q},
\label{eq:spe2}
\end{eqnarray}
and the curvature at the saddle point is found to be
\begin{eqnarray}
\hspace*{-15mm}
\Gamma(\bvarphi,\bvarphi^\prime)&=&
\frac{q(q\!-\!1)}{Z_q}\!\int\!\rmd\bpsi~\nu(\bpsi)U_1(\bvarphi,\bpsi)U_1(\bpsi,\bvarphi^\prime)
\Big[\int\!\rmd\bpsi^\prime~U_1(\bpsi,\bpsi^\prime)W_0(\bpsi^\prime)\Big]^{q-2}
\nonumber
\\[-0.5mm]
\hspace*{-15mm}
&& \hspace*{-3mm}
-qU_1(\bvarphi,\bvarphi^\prime)-\frac{q^2}{Z_q^2}\!
\int\!\rmd\bpsi~\nu(\bpsi)U_1(\bpsi,\bvarphi)
\Big[\int\!\rmd\bpsi^\prime~U_1(\bpsi,\bpsi^\prime)W_0(\bpsi^\prime)\Big]^{q-1}
\nonumber
\\[-0.5mm]
\hspace*{-15mm}
&&\hspace*{21.5mm}\times 
\int\!\rmd\bpsi~\nu(\bpsi)U_1(\bpsi,\bvarphi^\prime)
\Big[\int\!\rmd\bpsi^\prime~U_1(\bpsi,\bpsi^\prime)W_0(\bpsi^\prime)\Big]^{q-1}
\nonumber
\\
\hspace*{-15mm}
&=& q(q\!-\!1)\int\!\rmd\bpsi~U_1(\bvarphi,\bpsi)r[W_0(\bpsi)]U_1(\bpsi,\bvarphi^\prime)-qU_1(\bvarphi,\bvarphi^\prime)
\nonumber
\\[-0.5mm]
\hspace*{-15mm}&&\hspace*{10mm}-q^2\Big[\int\!\rmd\bpsi~U_1(\bvarphi,\bpsi)W_0(\bpsi)\Big]\Big[\int\!\rmd\bpsi~U_1(\bvarphi^\prime,\bpsi)W_0(\bpsi)\Big]
\nonumber
\\
\hspace*{-15mm}
&=& -q\int\!\rmd\bpsi~U_1(\bphi,\bpsi)\Big[\delta(\bpsi\!-\!\bvarphi^\prime)-T(\bpsi,\bvarphi^\prime)\Big],
\end{eqnarray}
with 
\begin{eqnarray}
\hspace*{-20mm}
T(\bvarphi,\bvarphi^\prime)=(q\!-\!1)r[W_0(\bvarphi)]U_1(\bvarphi,\bvarphi^\prime)-qW_0(\bvarphi)\int\!\rmd\bvarphi^\pprime U_1(\bvarphi^\prime,\bvarphi^\pprime)W_0(\bvarphi^\pprime).
\end{eqnarray}
We could also have written the curvature in the form $\bGamma=-q\bU^{\frac{1}{2}}(\one-\bT)\bU_1^{\frac{1}{2}}$ with 
a symmetric kernel $\bT$, but since we only require the determinant of $\bGamma$  this would not
make a difference. Various terms in $\phi[\hat{\varrho}]$  can be simplified using equations (\ref{eq:spe1},\ref{eq:spe2}). 
For instance, with the simple identity
 $\int\!\rmd\bvarphi\rmd\bvarphi^\prime W_0(\bvarphi)U_1(\bvarphi,\bvarphi^\prime)W_0(\bvarphi^\prime)=1$ we find that 
\begin{eqnarray}
\calS_0[W_0]&=& -\frac{q}{2}+\log Z_q,
\\[-0.2mm]
\calS_1[W_0]&=& \frac{1}{2}q^2-\frac{1}{4}(q\!-\!1)^2\int\!\rmd\bvarphi\rmd\bvarphi^\prime~r[W(\bvarphi)]U_2(\bvarphi,\bvarphi^\prime)r[W(\bvarphi^\prime)]
\nonumber
\\[-0.5mm]
&&\hspace*{20mm}-\frac{1}{2}(q\!-\!1)\int\!\rmd\bvarphi~r[W(\bvarphi)]V(\bvarphi).
\end{eqnarray}
Hence, using $\log{\rm Det}\bA={\rm Tr}\log\bA$ and apart from irrelevant additive constants:
\begin{eqnarray}
\hspace*{-15mm}
\phi[\hat{\varrho}]&=& \lim\Big\{
\log Z_q-\frac{(q\!-\!1)^2}{4N}\int\rmd\bvarphi\rmd\bvarphi^\prime~r[W_0(\bvarphi)]U_2(\bvarphi,\bvarphi^\prime)r[W_0(\bvarphi^\prime)]
\nonumber
\\
\hspace*{-15mm}
&&-\frac{q\!-\!1}{2N}\int\!\rmd\bvarphi~r[W_0(\bvarphi)]V(\bvarphi)-\frac{1}{2N}{\rm Tr}\log (\one\!-\!\bT)\Big\}+{\mathcal O}(N^{-\frac{3}{2}}).
\end{eqnarray}
We can expand the last nontrivial term using ${\rm Tr}\log(\one\!-\!\bT)=-\sum_{\ell=1}^\infty {\rm Tr}(\bT^\ell)/\ell$. The
first two terms in this sum give, after some simple manipulations:
\begin{eqnarray}
\hspace*{-10mm}
{\rm Tr}(\bT)&=& \int\!\rmd\bvarphi~T(\bvarphi,\bvarphi)~=~(q\!-\!)\int\!\rmd\bvarphi~r[W_0(\bvarphi)]V(\bvarphi)-q,
\\
\hspace*{-10mm}
{\rm Tr}(\bT^2)&=& \int\!\rmd\bvarphi\rmd\bvarphi^\prime~T(\bvarphi,\bvarphi^\prime)T(\bvarphi^\prime,\bvarphi)
\nonumber
\\
\hspace*{-10mm}
&=& 2q-q^2+(q\!-\!1)^2 \int\!\rmd\bvarphi\rmd\bvarphi^\prime~r[W_0(\bvarphi)]U_2(\bvarphi,\bvarphi^\prime)
r[W_0(\bvarphi^\prime)].
\end{eqnarray}
Thus the first two terms $\ell=1,2$ precisely remove those non-constant terms in $\phi[\hat{\varrho}]$ 
that originated from $\calS_1[W_0]$. This simplifies $\phi[\hat{\varrho}]$ to the following expression, modulo additive constants and ${\mathcal O}(N^{-\frac{3}{2}})$ terms, and upon inserting the definition of $Z_q$:\begin{eqnarray}
\hspace*{-20mm}
\phi[\hat{\varrho}]&=& \lim\Big\{
\log\int\!\rmd\bvarphi~\nu(\bvarphi)\Big[\int\!\rmd\bvarphi^\prime~U_1(\bvarphi,\bvarphi^\prime)W_0(\bvarphi^\prime)\Big]^q\!+\frac{1}{2N}\sum_{\ell=3}^\infty\frac{{\rm Tr}(\bT^\ell)}{\ell}\Big\}.
\end{eqnarray}

\section{Replica symmetric value of the traces}
\label{app:traces}

Here we compute the traces ${\rm Tr}(\bM^\ell)$, that appear in the generating function $\phi[\hat{\varrho}]$, for the kernel (\ref{eq:RS_M}). Upon defining $   \bmu^\star = \bmu  \!+\! (q\!-\!2)(\bX^*)^{-1}$ we can write this kernel as
\begin{eqnarray}
M(\bvarphi,\bvarphi^\prime)&=& \calZ^{-1}_{q-1}
\rme^{\frac{1}{2}\rmi\bvarphi\cdot\bmu^\star\bvarphi-\rmi\bvarphi\cdot\bvarphi^\prime}.
\end{eqnarray}
We can write the  $\ell$-th trace of $\bM$ as follows, with the identification $\bvarphi_{\ell+1}\equiv \bvarphi_1$, and that both $\bmu$ and $\bX^{\star}$ (and its inverse) are diagonal matrices in the space of $\bvarphi$:
\begin{eqnarray}
\hspace*{-17mm} 
    \Tr(\bM^\ell) & =& \calZ^{-\ell}_{q-1} \int\!\Big(\prod_{k=1}^\ell \rmd\bvarphi_k\Big)\Big( \prod_{k=1}^\ell  \rme^{
    \frac{1}{2}\rmi\bvarphi_k\cdot\bmu^\star\bvarphi_k-\rmi\bvarphi_k\cdot\bvarphi_{k+1}}
    \Big)
  \nonumber    \\
  \hspace*{-17mm} 
  & =& \calZ^{-\ell}_{q-1} \int\!\Big(\prod_{k=1}^\ell \rmd\bvarphi_k\Big) \rme^{
    \frac{1}{2}\rmi \sum_{k k^\prime=1}^\ell \bvarphi_k\cdot\big[\bmu^\star \delta_{kk^\prime}-(\delta_{k+1,k^\prime}+\delta_{k-1,k^\prime})\one \big]\bvarphi_{k^\prime}}
  \nonumber  
 \\
 \hspace*{-17mm} 
    &=&  \calZ^{-\ell}_{q-1} \prodmu \Bigg\{\proda \Big[\int \!\Big(\prod_{k=1}^\ell \rmd{\phi_k}\Big)
   \rme^{-\frac{\rmi}{2} \sum_{k k'=1}^\ell \phi_k \p{\delta_{k,k^\prime\! +1} + \delta_{k,k^\prime\!-1} -\delta_{kk^\prime}\mu^\star}\phi_{k^\prime}}\Big]
     \nonumber\\
     \hspace*{-17mm} 
     &&\hspace{13mm}\times \prodb \Big[\int\!\Big(\prod_{k=1}^\ell \rmd{\psi_k}\Big) 
   \rme^{\frac{\rmi}{2} \sum_{kk=1'}^\ell {\psi_k} \p{\delta_{k,k^\prime\!+1} + \delta_{k,k^\prime\!-1} -\delta_{kk^\prime}\overline{\mu^\star}}{\psi_{k'}}}\Big]
     \Bigg\}
     \nonumber\\
     \hspace*{-17mm} 
     &=&  \calZ^{-\ell}_{q-1} \prodmu \Big[Z(\mue|\bA^\star_{\ell,\mu})^{n_\mu}\overline{Z(\mue|\bA^\star_{\ell,\mu})}^{m_\mu}\Big].
\end{eqnarray}
Here $Z(\mue|\bA^\star_{\ell,\mu})$ denotes the original complex Gaussian integral defined in (\ref{eq:originalZ}), and  $\bA^\star_{\ell,\mu}$ is now the adjacency matrix of a loop of length $\ell$ in the presence of a complex field acting on  the diagonal, of value $(2\!-\!q)/\xmu$:
\begin{eqnarray}
    \p{\bA^\star_{\ell,\mu}}_{kk^\prime} &=& \delta_{k,k^\prime\!+1} +\delta_{k,k^\prime\!-1} +\frac{2\!-\!q}{\xmu}\delta_{kk^\prime}~~~~~({\rm with}~k~{\rm mod}~\ell).
\end{eqnarray}

\section{Recovering the Kesten-MacKay law}
\label{app:MacKay}

Here we derive expressions (\ref{eq:KMlaw}) and (\ref{eq:hdistribution}). The factor between curly brackets in the first line of  the generating function  (\ref{eqLgenerator_before_McKay}), which will give the eigenvalue spectrum for graphs from the ensemble (\ref{eq:ensemble}) in the limit $N\to\infty$, is given by the following expression, in which $x^\star(\mu)$ is given by (\ref{solutionXstar}):
\begin{eqnarray}
\hspace*{-21mm}
    \varrho_0 (\mu) 
&=&
 \frac{1}{2\pi} \dmu~{\rm Im}\Big[
    (q\!-\!2)
    \log \Big(x^\star(\mu)\!-\!\frac{1}{x^\star(\mu)}\Big)
-q  \log x^\star(\mu)\Big]
\nonumber
\\
\hspace*{-21mm}
&=&
 \frac{1}{2\pi} \dmu\Big[
    (q\!-\!2)
    {\rm Arg} \big(x^\star(\mu)\!-\!1\big)\!+ \! (q\!-\!2)
    {\rm Arg} \big(x^\star(\mu)\!+\!1\big)
\!-\!2(q\!-\!1)  {\rm Arg}\big(x^\star(\mu)\big)\Big].
\nonumber
\\[-0mm]
\hspace*{-21mm}
&&
\end{eqnarray}
We note that as soon as $\mu^2>4(q\!-\!1)$ we have $x^\star(\mu)\in{\rm I\!R}$, hence for all such eigenvalues  ${\rm Arg}\big(x^\star(\mu)\!+\!a\big)=0$ for any real $a$, and thus $\varrho_0(\mu)=0$. For eigenvalues $\mu^2<4(q\!-\!1)$, on the other hand, we may use identities such as 
  $\rmd{\rm Arg}(z)/\rmd\mu={\rm Im}(z^{-1}\rmd z/\rmd\mu)$ and $\rmd x^\star(\mu)/\rmd\mu=-\rmi x^\star(\mu)/\sqrt{4(q\!-\!1)\!-\!\mu^2}$ to derive:\begin{eqnarray}
\varrho_0(\mu)&=&
\frac{1}{2\pi}{\rm Im}\Big\{\frac{\rmd x^\star(\mu)}{\rmd\mu}
\Big[\frac{q-2}{x^\star(\mu)\!+\!1}
+\frac{q-2}{x^\star(\mu)\!-\!1}-\frac{2(q\!-\!1)}{x^\star(\mu)}\Big]\Big\}
\nonumber
\\
&=&
\frac{1}{\pi\sqrt{4(q\!-\!1)\!-\!\mu^2}}~{\rm Re}\Big\{
q\!-\!1-(q\!-\!2)\frac{x^{\star 2}(\mu)}{x^{\star 2}(\mu)\!-\!1}
\Big\}
\nonumber
\\[1mm]
&=&
\frac{q}{2\pi}\frac{\sqrt{4(q\!-\!1)\!-\!\mu^2}}{q^2-\mu^2}.
\end{eqnarray}
Hence, in combination, 
\begin{eqnarray}
\varrho_0(\mu)&=&
\frac{q}{2\pi}\frac{\sqrt{4(q\!-\!1)\!-\!\mu^2}}{q^2\!-\mu^2}~\theta\big[2\sqrt{q\!-\!1}\!-\!|\mu|\big].
\end{eqnarray}
In the same way we derive expression  (\ref{eq:hdistribution}) for the function $h(\mu)$:
\begin{eqnarray}
h(\mu)&=&
-\frac{1}{\pi}{\rm Im} \Big[\frac{-\rmi x^\star(\mu)}{\sqrt{4(q\!-\!1)\!-\!\mu^2}}\Big]
= \frac{1}{\pi}\frac{\theta\p{2\sqrt{q\!-\!1}-\abs{\mu}}}{\sqrt{4(q\!-\!1) \!-\! \mu^2 }}.
\end{eqnarray}

\section{Expected number of subgraphs in a RRG}
\label{app:subgraphsRRG}

Here we restate and apply  a result  in \cite{gao2008distribution} on the  expected number ${\rm E}[\bJ]$ of  strictly balanced subgraphs $\bJ$ with $k$ nodes and $\ell$ edges, in a random regular graph $\bA$ with $N$ nodes and degree $q$ (see \cite{gao2008distribution}  for the precise definition of strictly balanced subgraphs, here we only require that these include loops and cliques). This number is given by 
\begin{eqnarray}
\hspace*{-10mm}
    \textrm{E}[\bJ] = P(\bJ\subset \bA)\frac{[N]_k}{a(\bJ)},~~~~~~
    P(\bJ\subset \bA) = \frac{\prod_{i=1}^N [q]_{j_i}}{(qN)^\ell} \Big(1 \!+\! {\cal O}\big((\frac{qk}{N})^2\big)\Big)
\end{eqnarray}
Here $P(\bJ \subset \bA)$ is the probability of $\bJ$ being a subgraph of $\bA$, $j_i$ is the degree of node $i$ when computed only via incident links that belong to $\bJ$, $[r]_s=r!/(r\!-\!s)!$, and $a(\bJ)$ is the number of automorphisms of $\bJ$.
For the case of a length-$\ell$ loop, $\bJ=\bA_\ell$, we have $k=\ell$ and $a(\bJ) =2\ell$, and hence
\begin{eqnarray}
    \textrm{E}[\bA_\ell]=(q-1)^\ell/2 \ell
\end{eqnarray}
For $(q\!+\!1)$-node  cliques we have $k=q\!+\!1$, $\ell=\frac{1}{2}q(q\!+\!1)$, and $a(\bJ) = (q\!+\!1)!$, so 
\begin{eqnarray}
    \textrm{E}[K_{q+1}] = \frac{(q!)^{q+1}}{(Nq)^{\frac{q}{2}(q+1)}}\frac{[N]_{q+1}}{(q\!+\!1)!}
\end{eqnarray}

\end{document}